\documentstyle[11pt,amsmath,amssymb]{article}

\numberwithin{equation}{section}

\newcommand{\beq}{\begin{equation}}
\newcommand{\eeq}{\end{equation}}
\newcommand{\bex}{\begin{example}}
\newcommand{\eex}{\end{example}}
\newcommand{\ber}{\begin{remark}}
\newcommand{\eer}{\end{remark}}
\def\bel{\begin{lemma}}
\def\eel{\end{lemma}}
\def\bet{\begin{theorem}}
\def\eet{\end{theorem}}
\def\bed{\begin{definition}}
\def\eed{\end{definition}}
\def\bea{\begin{assumption}}
\def\eea{\end{assumption}}

\newcommand{\one}{\mathchoice {\rm 1\mskip-4mu l} {\rm 1\mskip-4mu l}
{\rm 1\mskip-4.5mu l} {\rm 1\mskip-5mu l}}
\newcommand{\cc}{{\mathbb C}}
\newcommand{\R}{{\mathbb R}}
\newcommand{\N}{{\mathbb N}}
\newcommand{\Z}{{\mathbb Z}}
\newcommand{\f}{{\rm f}}
\newcommand{\F}{{\rm F}}
\newcommand{\Tr}{{\rm Tr}}
\newcommand{\bpsi}{{\bar{\psi}}}
\newcommand{\fh}{{\frak h}}
\newcommand{\fg}{{\frak g}}

\newcommand{\fA}{{\frak A}}
\newcommand{\sinc}{{\rm sinc}}
\newcommand{\qed}{{$\blacksquare$}}
\newcommand{\eba}{{\emph{\bf a}}}
\newcommand{\ebb}{{\emph{\bf b}}}
\newcommand{\ebc}{{\emph{\bf c}}}
\newcommand{\ebd}{{\emph{\bf d}}}

\newcommand{\bbbone}{\mathchoice {\rm 1\mskip-4mu l} {\rm 1\mskip-4mu l}
{\rm 1\mskip-4.5mu l} {\rm 1\mskip-5mu l}}
\newcommand{\scalprod}[2]{\left\langle {#1}, {#2}\right\rangle}

\newcommand{\fer}[1]{(\ref{#1})}
\newcommand{\ran}{{\rm Ran\,}}

\newcommand{\h}{{\cal H}}
\newcommand{\cx}{{\mathbb C}}
\newcommand{\r}{{\mathbb R}}

\newcommand{\av}[1]{\left\langle{#1}\right\rangle}

\newcommand{\mm}{{\frak M}}

\newcommand{\spec}{{\rm spec}}
\newcommand{\e}{{\rm e}}

\newcommand{\ee}{{\cal E}}
\newcommand{\s}{{\cal S}}
\newcommand{\Lt}{{\tilde L}}
\newcommand{\C}{{\cal C}}

\renewcommand{\tilde}[1]{\widetilde{#1}}
\renewcommand{\i}{{\rm i}}

\renewcommand{\b}{{\cal B}}
\renewcommand{\d}{{\rm d}}
\renewcommand{\c}{{\cal C}}

\newcounter{resultcounter}[section]

\newtheorem{theorem}[resultcounter]{Theorem}
\newtheorem{lemma}[resultcounter]{Lemma}
\newtheorem{proposition}[resultcounter]{Proposition}

\begin{document}

\setcounter{page}{0}

\title{Asymptotics of repeated interaction \\
quantum systems}

\author{ 
Laurent Bruneau\footnote{
CPT-CNRS, UMR 6207
Universit\'e du Sud, Toulon-Var, BP 20132, 
F-83957 La Garde Cedex, France. Email: bruneau@cpt.univ-mrs.fr
}\ , Alain Joye\footnote{
Institut Fourier, UMR 5582, CNRS-Universit\'e de Grenoble I
BP 74, 38402 Saint-Martin d'H\`eres, France. Email: Alain.Joye@ujf-grenoble.fr
}\,\,\footnote{Laboratoire de Physique et Mod\'elisation des Milieux
  Condens\'es, UMR 5493, CNRS-Universit\'e de Grenoble I, BP 166,
  38042 Grenoble, France}\ , Marco Merkli\footnote{
Department of Mathematics and Statistics,
McGill University,
805 Sherbrooke Street West,
Montreal, QC, H3A 2K6
Canada. Email: merkli@math.mcgill.ca, http://www.math.mcgill.ca/\ $\widetilde{}$\,merkli/}
}
\date{\today}

\maketitle
\vspace{-1cm}
\begin{abstract}
A quantum system $\s$ interacts in a successive way with elements $\ee$ of a chain of identical independent quantum subsystems. Each interaction lasts for a duration $\tau$ and is governed by a fixed coupling between $\s$ and $\ee$. We show that the system, initially in any state close to a reference state, approaches a {\it repeated interaction asymptotic state} in the limit of large times. This state is $\tau$--periodic in time and does not depend on the initial state.  If the reference state is chosen so that $\s$ and $\ee$ are individually in equilibrium at positive temperatures, then the repeated interaction asymptotic state satisfies an average second law of thermodynamics.
\end{abstract}

\thispagestyle{empty}
%\newpage
\setcounter{page}{1}
\setcounter{section}{1}

\pagestyle{myheadings}
\markright{BJM \today}

\setcounter{section}{0}

\section{Introduction}

In this introduction we outline our main results and the relevant ideas of their proofs.

Suppose a quantum system $\s$ interacts with another one, $\ee$, during a time interval $[0,\tau)$, where $\tau>0$ is fixed. Then, for times $[\tau,2\tau)$, $\s$ interacts in the same fashion with another copy of $\ee$, and so on. The assembly of the systems $\ee$ (which are not directly coupled among each other) is called a chain, $\c=\ee +\ee +\cdots$. The system $\s+\c$, with an interaction as described above, is called a repeated interaction quantum system. One may think of $\s$ as being the system of interest, like a particle enclosed in a container, and of $\c$ as a chain of measuring apparatuses $\ee$ that are brought into contact with the particle in a sequential manner. 

Our goal is to study the large time behaviour of repeated interaction quantum systems, and in particular, to describe the effect of the repeated interaction on the system $\s$. One of our main results is the construction of the time--asymptotic state, which we call a {\it repeated interaction asymptotic state} (RIAS).

States of $\cal S$ and $\cal E$ are represented by vectors (or density matrices) in the Hilbert spaces $\h_\s$ and $\h_\ee$, respectively. We assume that $\dim\h_\s <\infty$, while $\dim\h_\ee\leq\infty$.  
The observables of $\s$ and $\ee$ are bounded operators, they form the (von Neumann) algebras $\mm_\s\subset{\cal B}(\h_\s)$ and $\mm_\ee\subset{\cal B}(\h_\ee)$. Observables evolve according to the Heisenberg dynamics $t\mapsto \tau_\s^t(A_\s)$ and $t\mapsto \tau_\ee^t(A_\ee)$, where $\tau_\s^t$ and $\tau_\ee^t$ are groups of $*$automorphisms of $\mm_\s$ and $\mm_\ee$, respectively.
  
We assume that there are distinguished ``reference'' states, represented by the vectors $\Omega_\s\in\h_\s$ and $\Omega_\ee\in\h_\ee$, and for the purposes of the introduction, we shall take $\Omega_\s$, $\Omega_\ee$ to be equilibrium states with respect to $\tau^t_\s$, $\tau^t_\ee$, for inverse temperatures $\beta_\s$, $\beta_\ee$, respectively. It is useful to pass to a description of the dynamics of vectors in $\h_\s$, $\h_\ee$ (Schr\"odinger dynamics). There are selfadjoint operators $L_\s$, $L_\ee$, called the standard Liouville operators, uniquely specified by 
\begin{equation}
\tau_\#^t(A)=\e^{\i tL_\#}A\e^{-\i tL_\#},\ \ \mbox{and}\ \  L_\#\Omega_\#=0,
\label{i1}
\end{equation}
where $\#$ stands here for either $\s$ or $\ee$. 

The Hilbert space of the entire system is given by $\h=\h_\s\otimes\h_\c$, where $\h_\c$, the Hilbert space of the chain, is the infinite tensor product $\otimes_{m\geq 1}\h_\ee$. The non-interacting dynamics is defined on the algebra $\mm_\s\otimes_{m\geq 1}\mm_\ee$ by $\tau^t_\s\otimes_{m\geq 1}\tau^t_\ee$. 

We consider interactions of the following kind. Fix an interaction time $\tau>0$. During the interval $[0,\tau)$, $\s$ interacts with the first element $\ee$ in the chain $\c$, while all other $\ee$'s evolve freely. The interaction is specified by an operator $V\in \mm_\s\otimes\mm_\ee$. In the next time interval, $[\tau,2\tau)$, $\s$ interacts with the second element in the chain, through the same interaction operator $V$, and all other elements evolve freely, and so on. For $t\geq 0$ we set
\begin{equation}
t=m(t)\tau +s(t),
\label{i2}
\end{equation}
where $m(t)$ is the integer measuring how many complete interactions of duration $\tau$ have taken place at the instant $t$, and where $0\leq s(t)<\tau$. We define the {\it repeated interaction (Schr\"odinger) dynamics}, for $t\geq 0$, $\psi\in\h$, by
\begin{equation}
U_{\rm RI}(t)\psi = \e^{-\i s(t) \Lt_{m(t)+1}} \e^{-\i\tau \Lt_{m(t)}}\cdots \e^{-\i\tau\Lt_1}\psi,
\label{i3}
\end{equation}
where
\begin{equation}
\Lt_m = L_m +\sum_{k\neq m} L_{\ee,k}
\label{i4}
\end{equation}
is the generator of the total dynamics during the time interval $[(m-1)\tau,m\tau)$. We have introduced $L_m$, the operator on $\h$ that acts trivially on all elements of the chain except for the $m$--th one, and which, on the remaining part of $\h$ (which is just $\h_\s\otimes\h_\ee$), acts as 
\begin{equation}
L=L_\s+L_\ee+V.
\label{i5}
\end{equation}
In \fer{i4}, $L_{\ee,k}$ denotes the operator on $\h$ that acts nontrivially only on the $k$--th element of the chain, on which it equals $L_\ee$. 

\medskip
A state $\omega$ given by a density matrix on $\h$ is said to be normal. Our goal is to understand the time-asymptotics ($t\rightarrow\infty$) of expectations
\begin{equation}
\omega\big(U_{\rm RI}(t)^* O U_{\rm RI}(t)\big)\equiv \omega\big(\alpha_{\rm RI}^t(O)\big),
\label{i6}
\end{equation}
for normal states $\omega$ and for certain classes of ``observables'' $O$. As mentioned above, we may regard $\s$ as the system of interest, so we certainly want to treat the case $O\in\mm_\s$. Another type of physical observable is of interest as well. Imagine we want to measure the variation, say, of the energy of $\s$ at a certain moment in time. This measuring process involves the system $\s$, but also the element of the chain that is in contact with $\s$ at the given moment. We call such an observable an {\it instantaneous observable}. Various generalizations can be considered, see Section \ref{modelresultssection}, but we limit our discussion in this introduction to the two kinds of observables just described. 

\medskip
{\bf Asymptotic state.\ } Let $O$ be an instantaneous observable, determined by $A_\s$ and $A_\ee$. This means that at time $t=m(t)\tau+s(t)$, \fer{i2}, $O$ measures $A_\s\otimes A_\ee$ on the system $\s +\ee$, where $\ee$ is the $(m(t)+1)$--th element in the chain $\c$. We show in Theorem \ref{thm1} that, under a natural assumption on the interaction, we have 
\begin{equation}
\Big| \omega\big(\alpha_{\rm RI}^t(O)\big) - \omega_+\big(P \alpha_{\rm RI}^{s(t)}(A_\s\otimes A_\ee)P\big)\Big|\longrightarrow 0,
\label{i7}
\end{equation}
as $t\rightarrow\infty$, where $\omega_+$ is a state on $\mm_\s$ which does not depend on $\omega$ (c.f. \fer{i9}), and where 
\begin{equation}
P=\bbbone_{\h_\s}\otimes_{m\geq 1}P_{\Omega_\ee}, 
\label{P}
\end{equation}
with $P_{\Omega_\ee}$ denoting the orthogonal projection onto $\cx\Omega_\ee$. We identify the range of $P$ with $\h_\s$. Relation \fer{i7} shows that the expectation of an instantaneous observable in any normal initial state approaches a $\tau$--periodic limit function ($t \mapsto s(t)$ is $\tau$--periodic). The speed of convergence in \fer{i7} is exponential, $\sim \e^{-t\gamma/\tau}$, where $\gamma>0$ is a constant depending on the interaction. 

The restriction of the RIAS to the algebra of instantaneous observables characterized by $A_\s\in\mm_\s$ and $A_\ee\in\mm_\ee$ is the $\tau$--periodic state 
\begin{equation}
A_\s\otimes A_\ee\mapsto \omega_+\big(P\alpha_{\rm RI}^{s(t)}(A_\s\otimes A_\ee)P\big)
\label{i7.1}
\end{equation}
on $\mm_\s\otimes\mm_\ee$, see \fer{i7} (and \fer{RIAS} for the definition of the RIAS acting on more general observables).

The above--mentioned assumption on the interaction is an ergodicity assumption on the dynamics reduced to the system $\s$. More precisely, we construct a (non--symmetric) operator $K$ on $\h$ s.t. 
\begin{equation}
P\alpha_{\rm RI}^t(A_\s)P \Omega_{\s}= (P\e^{\i \tau K}P)^{m(t)} P\e^{\i s(t)K}A_\s P\Omega_\s,
\label{i8}
\end{equation}
with the property 
\begin{equation}
P\e^{\i\tau K}P\Omega_\s=\Omega_\s. 
\label{i8.1}
\end{equation}
We assume that $1$ is a simple eigenvalue of $P\e^{\i\tau K}P$, and that all the other eigenvalues lie strictly inside the complex unit disk. (We prove in Section \ref{examplesection} that this holds for concrete models). As a consequence of this assumption, we have 
\begin{equation}
(P\e^{\i\tau K}P)^{m(t)}\longrightarrow |\Omega_\s\rangle\langle\Omega^*_\s|,
\label{i8.2}
\end{equation}
as $t\rightarrow\infty$, where $\Omega_\s^*$ is the unique vector in $\h_\s$ satisfying $(P\e^{\i \tau K}P)^*\Omega^*_\s=\Omega^*_\s$ and $\scalprod{\Omega^*_\s}{\Omega_\s}=1$. To arrive at result \fer{i7}, where
\begin{equation}
\omega_+(\cdot) =\scalprod{\Omega_\s^*}{\ \cdot\ \Omega_\s},
\label{i9}
\end{equation}
we use \fer{i8.2} together with an argument involving a cyclicity property of $\Omega_\s$. 

Our approach is constructive in the sense that the asymptotic characteristics of the system, such as the state $\omega_+$, the speed of convergence $\gamma$, and the asymptotic dynamics in \fer{i7} can be calculated by rigorous perturbation theory (in $V$).

\medskip
{\bf Correlations \& reconstruction of initial state.\ } The (time dependent) asymptotic expectations of observables do not depend on the initial state of the system, c.f. \fer{i7}. However, asymptotic correlations do, and together with the asymptotic expectations they permit to reconstruct the intial state in the following way.

Take a normal state $\omega$, an instantaneous observable $O$ determined by $A_\s\in\mm_\s$, $A_\ee\in\mm_\ee$, and an observable $A\in\mm$. We show in Theorem \ref{thm3} that 
\begin{equation}
\big|\omega\big( A \alpha_{\rm RI}^t(O)\big) -\omega(A)\, \omega_+\big(P\alpha_{\rm RI}^{s(t)}(A_\s\otimes A_\ee)P\big)\big|\longrightarrow 0,
\label{i10}
\end{equation}
as $t\rightarrow\infty$ (exponentially fast). According to \fer{i7} and \fer{i10}, knowledge of the asymptotic correlation function ${\cal C}_+(t;A,O)$, and of the asymptotic expectation $E_+(t;O)$, determined respectively by 
\begin{equation*}
\lim_{t\rightarrow\infty}|\omega(A\alpha_{\rm RI}^t(O))- {\cal C}_+(t;A,O)|=0 \mbox{\ \ and\ \ } \lim_{t\rightarrow\infty}|\omega(\alpha_{\rm RI}^t(O))-E_+(t;O)|=0, 
\end{equation*}
allows for a reconstruction of the initial state $\omega$ according to 
\begin{equation}
\omega(A) = \frac{{\cal C}_+(t;A,O)}{E_+(t;O)}.
\label{i11}
\end{equation}

\medskip
{\bf Energy, entropy, average $\bf 2^{\rm \bf nd}$ law of thermodynamics for RIAS. } The formal quantity $\alpha_{\rm RI}^t(\widetilde L_{m(t)+1})$, where $\widetilde L_m$ is given by \fer{i4}, has a well-defined variation in $t$. It is not hard to see by explicit calculation that this variation is zero in all time intervals $[(m-1)\tau,m\tau)$, and that it undergoes a jump 
\begin{equation*}
j(m)=\alpha_{\rm RI}^{m\tau}(V_{m+1}-V_m)
\end{equation*}
as time passes the moment $m\tau$. Here, $V_k$ denotes the operator $V$ acting nontrivially only on $\h_\s$ and the $k$--th element $\h_\ee$ of the chain Hilbert space $\h_\c$. We interpret the variation of the above formal quantity as the (time dependent) observable of variation in total energy of the system. 

We show in Section \ref{energysubsection} that for any normal initial state $\omega$, the variation in energy during any time interval of length $\tau$ takes the asymptotic expectation value $\omega_+(j_+)$, where
\begin{equation}
j_+=PVP -P\alpha_{\rm RI}^\tau(V)P = -\i\int_0^\tau P\alpha_{\rm RI}^s([L_\s+L_\ee,V])P\d s.
\label{i12}
\end{equation}
(Here and in the rest of the paper we understand commutators to be defined in the form sense, but none of our arguments involve delicate domain questions with regards to commutators.) We define the (average) {\it asymptotic energy production $\d E_+$} to be the change in energy during any interval of duration $\tau$, divided by $\tau$, in the limit of large times. This quantity is given by 
\begin{equation}
\d E_+=\frac 1\tau \omega_+(j_+)
\label{i13}
\end{equation}
and is independent of the initial state $\omega$. We show in Section \ref{subsectionentropy} that $\omega_+(j_+)\geq 0$.

Denote by $\omega_0$ the state on $\mm$ determined by the vector $\Omega_\s\otimes_{m\geq 1}\Omega_\ee\in \h$, and let ${\rm Ent}(\omega|\omega_0)$ denote the relative entropy of the normal state $\omega$ w.r.t. $\omega_0$.
 We think it is natural to define the entropy as a non--negative quantity, and our definition of it differs by a sign from the one given in \cite{BR}\footnote{For a finite system we have ${\rm Ent}(\omega|\omega_0) = {\rm Tr}(\rho(\log\rho-\log\rho_0))$, where $\rho$ and $\rho_0$ are density matrices determining the states $\omega$ and $\omega_0$, respectively, and where $\rho_0>0$.}. 
We define the (average) {\it asymptotic entropy production $\d S_+$} to be the change of (relative) entropy in any interval of duration $\tau$, divided by $\tau$, in the limit of large times. We prove in Section \ref{subsectionentropy} that 
\begin{equation}
\d S_+ = \frac{\beta_\ee}{\tau} \omega_+(j_+),
\label{i14}
\end{equation}
where $\beta_\ee$ is the inverse temperature of the elements in the chain. 
The asymptotic entropy production does not depend on the initial state $\omega$. We may combine \fer{i13} and \fer{i14} to arrive at an {\it average {\rm 2$^{{\rm nd}}$} law of thermodynamics for repeated interaction quantum systems}, 
\begin{equation}
\d E_+= T_\ee\, \d S_+,
\label{i15}
\end{equation}
where $T_\ee=1/\beta_\ee$ is the temperature of the chain. Relation \fer{i15} is independent of the initial state of the system, and it holds for any repeated interaction system ($V$ and $\tau$).

\bigskip

Repeated interaction quantum systems emerge in quantum optics, see e.g. \cite{VAS,WBKM} and references therein. The effective evolution of the small system in certain specific regimes of parameters related to the Van Hove limit has been studied in \cite{AJ}. In some parameter regimes repeated interaction models can be considered as coarse grained versions of a system $\s$ in contact with quantum noises \cite{AP}.

\section{Model \& Results}
\label{modelresultssection}

\subsection{Repeated interaction models} 

The models we consider consist of a system $\cal S$ which is coupled to a chain $\c=\ee+\ee+\cdots$ of identical elements $\ee$. We describe $\s$ and $\ee$ as $W^*$--dynamical systems $(\mm_\s,\tau_\s^t)$ and $(\mm_\ee,\tau_\ee^t)$, where $\mm_\s$, $\mm_\ee$ are von Neumann algebras ``of observables'' acting on the Hilbert spaces $\h_\s$, $\h_\ee$, respectively, and where $\tau^t_\s$ and $\tau^t_\ee$ are ($\sigma$--weakly continuous) groups of $*$automorphisms describing the Heisenberg dynamics. In this paper, we consider the situation $\dim\h_\s<\infty$ and $\dim\h_\ee\leq \infty$.

We assume that there are distinguished vectors $\Omega_\s\in\h_\s$ and $\Omega_\ee\in\h_\ee$, determining states on $\mm_\s$ and $\mm_\ee$ which are invariant w.r.t. $\tau_\s^t$ and $\tau_\ee^t$, respectively, and we assume that $\Omega_\s$ and $\Omega_\ee$ are cyclic and separating for $\mm_\s$ and $\mm_\ee$, respectively. One may typically think of these distinguished vectors as being KMS vectors.

The Hilbert space of the chain $\c$ is defined to be the infinite tensor product \begin{equation}
\h_\c=\otimes_{m\geq 1}\h_\ee
\label{m1}
\end{equation}
w.r.t. the reference vector
\begin{equation}
\Omega_\c=\Omega_\ee\otimes\Omega_\ee\cdots.
\label{m16}
\end{equation}
In other words, $\h_\c$ is obtained by taking the completion of the vector space of finite linear combinations of the form $\otimes_{m\geq 1}\psi_m$, where $\psi_m\in\h_\ee$, $\psi_m=\Omega_\ee$ except for finitely many indices, in the norm induced by the inner product
\begin{equation}
\scalprod{\otimes_m\psi_m}{\otimes_m\chi_m} = {\textstyle \prod}_m\scalprod{\psi_m}{\chi_m}_{\h_\ee}.
\label{m2}
\end{equation}
We introduce the von Neumann algebra 
\begin{equation}
\mm_\c =\otimes_{m\geq 1}\mm_\ee
\label{m3}
\end{equation}
acting on $\otimes_{m\geq 1}\h_\ee$, which is obtained by taking the weak closure of finite linear combinations of operators $\otimes_{m\geq 1} A_m$, where $A_m\in\mm_\ee$ and $A_m=\bbbone_{\h_\ee}$ except for finitely many indices. 

The operator algebra containing the observables of the total system is the von Neumann algebra 
\begin{equation}
\mm = \mm_\s\otimes\mm_\c
\label{m4}
\end{equation}
which acts on the Hilbert space
\begin{equation}
\h=\h_\s\otimes\h_\c.
\label{m5}
\end{equation}

The {\it repeated interaction dynamics} of observables in $\mm$ is characterized by an interaction time $0<\tau<\infty$ and a selfadjoint  interaction operator \begin{equation}
V\in \mm_\s\otimes\mm_\ee. 
\label{mV}
\end{equation}
For times $t\in [\tau(m-1), \tau m)$, where $m\geq 1$, $\s$ interacts with the $m$--th element of the chain, while all other elements of the chain evolve freely (each one according to the dynamics $\tau_\ee$). 
The interaction of $\s$ with every element in the chain is the same (given by $V$). 

Let $L_\s$ and $L_\ee$ be the {\it standard Liouville operators} (``positive temperature Hamiltonians'', c.f. references of \cite{JP,MMS}), uniquely characterized by the following properties: $L_{\#}$ (where $\#=\s,\ee$) are selfadjoint operators on $\h_{\#}$ which implement the dynamics $\tau^t_{\#}$, 
\begin{equation}
\tau^t_{\#}(A) = \e^{\i t L_{\#}} A \e^{-\i tL_{\#}},\ \ \ \forall A\in\mm_{\#},
\label{m6}
\end{equation}
and 
\begin{equation}
L_{\#}\Omega_{\#}=0.
\label{m7}
\end{equation}
We define the selfadjoint operator
\begin{equation}
L = L_\s+L_\ee +V,
\label{m7.1}
\end{equation}
omitting trivial factors $\bbbone_\s$ or $\bbbone_\ee$ (by $L_\s$ in \fer{m7.1} we really mean $L_\s\otimes\bbbone_\ee$, etc). $L$ generates the automorphism group $\e^{\i t L}\cdot \e^{-\i tL}$ of $\mm_\s\otimes\mm_\ee$, the interacting dynamics between $\s$ and an element $\ee$ of the chain $\c$. The explicit form of the operator $V$ is dictated by the underlying physics, we give some examples in Section \ref{examplesection}.

For $m\geq 1$ let us denote by
\begin{equation}
\Lt_m = L_m +\sum_{k\neq m} L_{\ee,k}
\label{m8}
\end{equation}
the generator of the total dynamics during the interval $[(m-1)\tau,m\tau)$. We have introduced $L_m$, the operator on $\h$ that acts trivially on all elements of the chain except for the $m$--th one. On the remaining part of the space (which is isomorphic to $\h_\s\otimes\h_\ee$), $L_m$ acts as $L$, \fer{m7.1}. We have also set $L_{\ee,k}$ to be the operator on $\h$ that acts nontrivially only on the $k$--th element of the chain, on which it equals $L_\ee$. Of course, the infinite sum in \fer{m8} must be interpreted in the strong sense on $\h$.

Decompose $t\in\r_+$ as 
\begin{equation}
t = m(t) \tau +s(t),
\label{m9}
\end{equation}
where $m(t)$ is the integer measuring the number of complete interactions of duration $\tau$ the system $\s$ has undergone at time $t$, and where $0\leq s(t)<\tau$.  The repeated interaction dynamics of an operator $A$ on $\h$ is defined by
\begin{equation}
\alpha^t_{\rm RI}(A) = U_{\rm RI}(t)^*  \,A \,U_{\rm RI}(t)
\label{m10}
\end{equation}
where 
\begin{equation}
U_{\rm RI}(t) = \e^{-\i s(t) \Lt_{m(t)+1}} \e^{-\i\tau \Lt_{m(t)}}\cdots \e^{-\i\tau\Lt_1}
\label{m11}
\end{equation}
defines the Schr\"odinger dynamics on $\h$. According to this dynamics $\s$ interacts in succession, for a fixed duration $\tau$ and a fixed interaction $V$, with the first $m(t)$ elements of the chain, and for the remaining duration $s(t)$ with the $(m(t)+1)$--th element of the chain. Being the propagator of a ``time-dependent Hamiltonian'' (which is piecewise constant), $U_{\rm RI}(t)$ does not have the group property in $t$.

\bigskip
Our goal is to examine the large time behaviour of expectation values of certain observables in normal states $\omega$ on $\mm$ (states given by a density matrix on $\h$). The system $\s$ feels an {\it effective dynamics} induced by the interaction with the chain $\c$. Under a suitable {\it ergodicity assumption} on this effective dynamics the small system is driven to an asymptotic state, as time increases. We will express the effective dynamics and the ergodic assumption using the modular data of the pair $(\mm_\s\otimes\mm_\c, \Omega_\s\otimes\Omega_\c)$. 

Let $J$ and $\Delta$ denote the modular conjugation and the modular operator associated to $(\mm_\s\otimes\mm_\ee, \Omega_\s\otimes\Omega_\ee)$, \cite{BR}. We assume that 
\begin{itemize}
\item[\bf (A)\ ] $\Delta^{1/2} V \Delta^{-1/2}\in \mm_\s\otimes\mm_\ee$
\end{itemize}
and we introduce the operator
\begin{equation}
K= L - J\Delta^{1/2} V\Delta^{-1/2} J,
\label{m14}
\end{equation}
called a $C$--Liouville operator, \cite{JP,MMS}. It generates a strongly continuous group of bounded operators, denoted $\e^{\i t K}$, satisfying $\|\e^{\i t K}\|\leq \e^{|t|\, \|\Delta^{1/2} V\Delta^{-1/2}\|}$. The main feature of the operator $K$ is that $\e^{\i tK}$ implements the same dynamics as $\e^{\i tL}$ on $\mm_\s \otimes\mm_\ee$ (since the difference $K-L$ belongs to the commutant $\mm_\s'\otimes\mm_\ee'$), and that 
\begin{equation}
K\Omega_\s\otimes\Omega_\ee=0.
\label{m15}
\end{equation}
Relation \fer{m15} follows from assumption (A), definition \fer{m14} and the properties $\Delta^{-1/2}J=J\Delta^{1/2}$  and $J\Delta^{1/2}A \Omega_\s\otimes\Omega_\ee= A^*\Omega_\s\otimes\Omega_\ee$, for any $A\in\mm_\s\otimes\mm_\ee$. 

Let 
\begin{equation}
P=\bbbone_{\h_\s}\otimes|\Omega_\c\rangle\langle\Omega_\c|
\label{mP}
\end{equation}
be the orthogonal projection onto $\h_\s\otimes\cx\Omega_\c\cong \h_\s$, where $\Omega_\c$ is given in \fer{m16}.  
If $B$ is an operator acting on $\h$ then we identify $PBP$ as an operator acting on $\h_\s$. We have 
\begin{proposition}
\label{proposition1}
There is a constant $C<\infty$ s.t. $\|(P\e^{\i t K}P)^m\|_{\b(\h_\s)}\leq C$, for all $t\in\r$, $m\geq 0$. In particular, $\spec(P\e^{\i tK}P)\subset \{z\in\cx\ |\ |z|\leq 1\}$ and all eigenvalues lying on the unit circle are semisimple.
\end{proposition}
We give a proof of Proposition \ref{proposition1} in Section \ref{proofprop1}. 
Relation \fer{m15} implies that for all $t\in\r$, $P\e^{\i t K}P\Omega_\s=\Omega_\s$.
Our assumption (E) on the effectiveness of the coupling is an ergodicity assumption on the discrete dynamics generated by 
\begin{equation}
M\equiv M(\tau)=P\e^{\i \tau K}P
\label{m17}
\end{equation}
\begin{itemize}
\item[\bf (E)] The spectrum of $M$ on the complex unit circle consists of the single eigenvalue $\{1\}$. This eigenvalue is simple (with corresponding eigenvector $\Omega_\s$).
\end{itemize}
Assumption (E) guarantees that the adjoint operator $M^*$ has a unique invariant vector, called $\Omega_\s^*$ (normalized as $\scalprod{\Omega_\s^*}{\Omega_\s}=1$), and that 
\begin{equation}
\lim_{m\rightarrow\infty} M^m = \pi:= |\Omega_\s\rangle\langle \Omega_\s^*|,
\label{m17.1}
\end{equation}
in the operator sense, where $\pi$ is the rank one projection which projects onto $\cx\Omega_\s$ along $(\cx\Omega_\s^*)^\perp$.  In fact, we have the following easy estimate (valid for any matrix $M$ with spectrum inside the unit disk and satisfying (E))
\begin{proposition}
\label{proposition3}
For any $\epsilon>0$ there exists a constant $C_\epsilon$ s.t. $\|M^m-\pi\|\leq C_\epsilon \e^{-m(\gamma-\epsilon)}$, for all $m\geq 0$, where $\gamma:=\min_{z\in{\rm spec}(M)\backslash\{1\}}|\log|z|\,|>0$.
\end{proposition}
The parameter $\gamma$ measures the speed of convergence.

{\it Remark.\ } If all eigenvalues of $M$ are semisimple then in Proposition \ref{proposition3} we have $\|M^m-\pi\|\leq C\e^{-m\gamma}$ for some constant $C$ and all $m\geq 0$.

As a last preparation towards an understanding of our results we discuss the kinds of observables we consider.  
One interesting such class is $\mm_\s\subset \mm$ which consists of observables of the system $\s$ only. There are other observables of interest. 
We may think of the system $\s$ as being fixed in space and of the chain as passing by $\s$ so that at the moment $t$, the $(m(t)+1)$--th element $\ee$ is located near $\s$, c.f. \fer{m9}. A detector placed in the vicinity of $\s$ can measure at this moment in time observables of $\s$ and those of the $(m(t)+1)$--th element in the chain, i.e., an ``instantaneous observable'' of the form $A_\s\otimes \vartheta_{m(t)+1}(B_0)$, where $A_\s\in\mm_\s$, $B_0\in\mm_\ee$, and  $\vartheta_m:\mm_\ee\rightarrow\mm_\c$ is defined by 
\begin{equation}
\vartheta_m(A_\ee) =\bbbone_\ee\cdots\bbbone_\ee\otimes A_\ee\otimes\bbbone_\ee\cdots
\label{m12}
\end{equation}
where the $A_\ee$ on the right side of \fer{m12} acts on the $m$--th factor in the chain. An example of such an observable is the energy flux (variation) of the system $\s$. More generally we may be interested in the expectation value of operators of the form
\begin{equation}
[A_\s;A_i;B_j]\equiv A_\s\otimes_{i=1}^p A_i\otimes_{j=-\ell}^r\vartheta_{m(t)+j+1}(B_j),
\label{mop}
\end{equation}
where $A_\s\in\mm_\s$, $A_1,\ldots,A_p\in\mm_\ee$, $B_{-\ell},\ldots, B_0, \ldots, B_r\in\mm_\ee$ and where $t= m(t) \tau +s(t)$ as in \fer{m9} and $\vartheta_k$ is given in \fer{m12}. The parameters $p\geq 1$, $\ell, r\geq 0$ are not displayed in the l.h.s. in \fer{mop}. (We always assume that $p<m(t)-\ell+1$.)
$A_\s$ and the $A_i$ represent observables we measure on the small system and on the element with index $i$ of the chain, the $B_0$ is the ``instantaneous'' observable, measured in the element $m(t)+1$ of the chain (the one in contact with $\s$ at time $t$), while the $B_j$ with negative and positive index are the quantities measured in the elements preceding and following the $(m(t)+1)$--th.

\subsection{Asymptotic state}

Throughout the paper we assume that Conditions (A) and (E) of the previous section are satisfied. 

We consider the large time limit of expectations
\begin{equation}
E(t)= \omega\big(\alpha_{\rm RI}^t [A_\s;A_i;B_j]\big)
\label{expval}
\end{equation}
for observables $[A_\s;A_i;B_j]$ as in \fer{mop} and for normal states $\omega$ on $\mm$. Define the state $\omega_+$ on $\mm_\s$ by
\begin{equation}
\omega_+(A_\s) = \scalprod{\Omega_\s^*}{A_\s \Omega_\s},
\label{m22}
\end{equation}
where $\Omega_\s^*$ is defined in \fer{m17.1}.

\begin{theorem}
\label{thm1}
Let $\omega$ be fixed and take $A_i=\bbbone_\ee$, $i=1,\ldots,p$. For any $\epsilon>0$ there is a constant $C_\epsilon$ s.t. for all $t\geq 0$
\begin{equation}
\left| E(t)-E_+(t)\right|\leq C_\epsilon\, \e^{-t(\gamma-\epsilon)/\tau},
\label{m-4}
\end{equation}
where $\gamma>0$ is given in Proposition \ref{proposition3}, and where $E_+$ is the $\tau$-periodic function 
\begin{equation}
E_+(t)=\omega_+\Big( P \alpha_{\rm RI}^{\ell\tau +s(t)}\Big(A_\s\otimes B_{-\ell}\otimes\cdots\otimes B_0 \Big)P\Big) {\textstyle \prod_{j=1}^r}\av{B_j}_{\Omega_\ee}.
\label{m-2} 
\end{equation}
Here, $\av{B_j}_{\Omega_\ee}=\scalprod{\Omega_\ee}{B_j\Omega_\ee}$. 
\end{theorem}

We define the RIAS to be the $\tau$--periodic state on $\mm_\s\otimes_{j=-\ell}^r\mm_\ee$ given by 
\begin{equation}
A_\s \otimes_{j=-\ell}^r B_j \mapsto \omega_+\Big( P \alpha_{\rm RI}^{\ell\tau +s(t)}\Big(A_\s\otimes B_{-\ell}\otimes\cdots\otimes B_0 \Big)P\Big) {\textstyle \prod_{j=1}^r}\av{B_j}_{\Omega_\ee}.
\label{RIAS}
\end{equation}
Using \fer{m-4} and the unicity of the limit, one can see
that actually the state $\omega_+$ \emph{does not} depend on the choice of the
reference state $\Omega_S$.

{\it Remarks.\ } 1)\ If $B_{-\ell},\ldots,B_{-\ell'-1}=\bbbone_{\ee}$ for some $-\ell'-1\leq -1$, then one shows that 
\begin{equation*}
\omega_+\Big(P \alpha_{\rm RI}^{\ell\tau +s(t)}\Big(A_\s\otimes B_{-\ell}\otimes\cdots\otimes B_0 \Big)P\Big)=
\omega_+\Big( P \alpha_{\rm RI}^{\ell'\tau +s(t)}\Big(A_\s\otimes B_{-\ell'}\otimes\cdots\otimes B_0 \Big)P\Big),
\end{equation*}
and in case $B_j=\bbbone_\ee$ for all $j=-\ell,\ldots, 0$ formula \fer{m-2} is understood with $\alpha_{\rm RI}^{\ell\tau+s(t)}$ replaced by $\alpha_{\rm RI}^{s(t)}$. 

2)\ $C_\epsilon$ in Theorem \ref{thm1} is uniform in $\tau$ for $\tau>0$ varying in compact sets, and it is uniform in  
 $\big\{A_\s\in\mm_\s, \{B_j\}_{j=1}^r \subset \mm_\ee\ \big|\ \|A_\s\|\, {\textstyle \prod_{j=1}^r} \|B_j\|\leq {\rm const.}\big\}$. 

3) The convergence is determined by that of Proposition
\ref{proposition3}. If the ergodic assumption (E) is not satisfied then the
limit $\lim_{n\to\infty} M^n$ still exists, in a
weaker sense. Namely, if there are eigenvalues different from $1$ on
the circle, then the limit exists in the ergodic mean sense, $\frac{1}{N}\sum_{n=0}^{N-1} M^n=\pi+O(\frac{1}{N})$.
Further, if $1$ is a degenerate eigenvalue of $M$ then the limit exists but the projection $\pi$ is not one dimensional. This reflects in Theorem \ref{thm1} in the following way. If  $1$ is non degenerate, but there are other eigenvalues on the circle, then Theorem \ref{thm1} holds with
\fer{m-4} replaced by
\beq
\left|\frac{1}{t}\sum_{m=0}^{m(t)}
  E(m(t)\tau+s(t))-\frac{E_+(t)}{\tau}\right|\leq \frac{C}{t}.
\eeq 
If on the other hand $1$ is degenerate but 
there is no other eigenvalue on the circle, then one can still prove that the expectation value $E(t)$ has
an aymptotic behaviour $E_\infty(t,\omega)$, which is $\tau-$periodic,
but which will a priori depend on the initial state $\omega$
(c.f. \fer{m52.1} in the proof of Theorem \ref{thm1}).  
Of course if both $1$ is degenerate and there are other eigenvalues on
the circle, then
one gets convergence to $E_\infty(t,\omega)$ but in the ergodic mean.

\medskip
Our next result incorporates the measurement of observables $A_1,\ldots A_p\in\mm_\ee$ for a chain consisting of {\it dispersive} systems $\ee$. We measure dispersivity by the property of return to equilibrium. $\ee$ is said to have the latter property iff for any normal state $\omega_\ee$ on $\mm_\ee$ we have the relation
\begin{equation}
\lim_{t\rightarrow\infty} \omega_\ee(\tau_\ee^t(A_\ee)) = \scalprod{\Omega_\ee}{A_\ee\Omega_\ee},
\label{mrte}
\end{equation}
for any $A_\ee\in\mm_\ee$. Examples of such $\ee$ include reservoirs of ideal quantum gases. It is worthwile to mention that $\ee$ has the property of return to equilibrium if and only if $\e^{\i tL_\ee}$ converges in the weak sense to the orthogal projection onto $\cx\Omega_\ee$, as $t\rightarrow\infty$.
\begin{theorem}
\label{thm2} 
Suppose $\ee$ has the property of return to equilibrium. 
Then
\begin{equation}
\lim_{t\rightarrow\infty}|E(t)-E_+(t)|=0,
\label{m-3.0}
\end{equation}
where $E_+(t)$ is the $\tau$-periodic function 
\begin{equation}
E_+(t) = \omega_+\Big( P \alpha_{\rm RI}^{\ell\tau +s(t)}\Big(A_\s\otimes  B_{-\ell}\otimes\cdots\otimes B_0 \Big)P\Big) {\textstyle \prod_{i=1}^p}\av{A_i}_{\Omega_\ee}{\textstyle \prod_{j=1}^r}\av{B_j}_{\Omega_\ee}.
\label{m-3}
\end{equation}
\end{theorem}

{\it Remark.\ } The speed of convergence in \fer{m-3.0} is determined by that of return to equilibrium, \fer{mrte}, and by $\gamma$, Proposition \ref{proposition3}. The limit \fer{m-3.0} is uniform in $\tau$, for $\tau$ varying in compact sets, and it is uniform in balls of observables $\|A_\s\|\, \prod_{i=1}^p\|A_i\|\prod_{j=-\ell}^r\|B_j\|\leq {\rm const.}$

\subsection{Correlations \& reconstruction of initial state}

As Theorems \ref{thm1} and \ref{thm2} show, the limiting expectation values $E_+(t)$ are independent of the initial state (the state $\omega_+$ is, c.f. \fer{m22}). However, limiting correlations are not, and their knowledge allows to reconstruct the initial state.

Fix a normal initial state $\omega$ of $\mm$ and let $A\in\mm$, $A_\s\in\mm_\s$, $B_0\in\mm_\ee$. We define the correlation between $A$ and the instantaneous observable $A_\s\otimes \vartheta_{m(t)+1}(B_0)$ by 
\begin{equation}
\C(t;A,A_\s,B_0) = \omega\left( A\,\alpha_{\rm RI}^t(A_\s\otimes\vartheta_{m(t)+1}(B_0))\right).
\label{m60}
\end{equation}
\begin{theorem}
\label{thm3}
For any $\epsilon>0$ there is a constant $C_\epsilon$ s.t. for all $t\geq 0$
\begin{equation}
\left| \,\C(t;A,A_\s,B_0) - \C_+(t;A,A_\s,B_0)\right|\leq C_\epsilon \e^{-t(\gamma-\epsilon)/\tau},
\label{m61}
\end{equation}
where $\gamma$ is given in Proposition \ref{proposition3}, and where $\C_+$ is the $\tau$--periodic limiting correlation function
\begin{equation}
\C_+(t;A,A_\s,B_0) = \omega(A)\  \omega_+\!\left(P\alpha_{\rm RI}^{s(t)}(A_\s\otimes B_0)P\right),
\label{m62}
\end{equation}
with $\omega_+$ defined in \fer{m17.1}. 
\end{theorem}

{\it Remark.\ } Relation \fer{m62} allows us to reconstruct the initial state $\omega$, knowing the asymptotic state $\omega_+$ and the asymptotic correlation function $\C_+$.

\subsection{Energy, entropy and their relation}

It may not be meaningful to speak about the total energy of the system, because it may have to be considered as being infinite, e.g. if the elements $\ee$ of the chain are infinitely extended quantum systems with non-vanishing energy density. However, we can define the time variation of the total energy of the system and link it to its entropy variation, giving us an average 2$^{\rm nd}$ law of thermodynamics for RIAS.

\subsubsection{Energy}
\label{energysubsection}

Recall that $\tilde L_{m+1}$, $m\geq 0$, is the generator of the total dynamics in the time interval $t=m\tau +s \in[m\tau,(m+1)\tau)$, during which the $(m+1)$--th element of the chain interacts with $\s$, c.f. \fer{m8}. Given any integer $m\geq 0$ and any $0\leq s<\tau$ it is easy to formally verify the relation
\begin{equation}
\alpha_{\rm RI}^{m\tau+s}(\tilde L_{m+1}) -\alpha_{\rm RI}^{m\tau}(\tilde L_{m+1}) =0.
\label{m67}
\end{equation}
This suggests that the formal quantity $\alpha_{\rm RI}^{t}(\tilde L_{m+1})$ is constant for $t$ in any interval $[m\tau,(m+1)\tau)$. Another short calculation yields that this quantity undergoes a jump $j(k)$ as time passes the moment $k\tau$, $k\geq 1$: for $(k-1)\tau\leq t_1< k \tau\leq t_2<(k+1)\tau$ we have 
\begin{equation}
j(k):=\alpha_{\rm RI}^{t_2}(\tilde L_{k+1}) -\alpha_{\rm RI}^{t_1}(\tilde L_{k})= \alpha_{\rm RI}^{k \tau}\big(V_{k+1}-V_k\big),
\label{m68}
\end{equation}
where we set 
\begin{equation}
V_k=[\bbbone_{\mm_\s}\otimes \vartheta_k](V)
\label{m68.1}
\end{equation}
(see \fer{m12}). We interpret $j(k)$ as the change in total energy as time passes the moment $k\tau$. 

Theorem \ref{thm1} tells us that for any normal state $\omega$ on $\mm$ and for any $\epsilon>0$, there is a constant $C_\epsilon$ s.t. 
\begin{equation}
\left| \omega(j(k))-\omega_+\big(j_+\big)\right| \leq C_\epsilon \e^{-k(\gamma-\epsilon)},
\label{m69}
\end{equation}
where
\begin{equation}
j_+= PVP -P\alpha_{\rm RI}^\tau(V)P = -\i \int_0^\tau P\alpha_{\rm RI}^s\big([L_\s+L_\ee,V]\big)P \, \d s.
\label{m70}
\end{equation}
Relation \fer{m69} and the fact that the energy is piecewise constant shows that $\omega_+(j_+)$ is the change of energy in any interval of length $\tau$, in the large time limit. We thus call 
\begin{equation}
\d E_+= \frac{1}{\tau} \omega_+(j_+)
\label{m82}
\end{equation}
the {\it asymptotic energy production}. The asymptotic energy production does not depend on the initial state of the system.

{\it Remark.\ } It is not hard to see that the expectation of the energy jump is constant in the state $\omega_+\otimes\omega_\c$, where $\omega_\c$ is the vector state on $\mm_\c$ determined by $\Omega_\c$, \fer{m16}:
\begin{equation}
\omega_+\otimes\omega_\c(j(k)) = \omega_+(j_+),\ \ \ \forall k\geq 1.
\label{m71}
\end{equation}

\medskip
We introduce the variation of the total energy, $\Delta E(t)$, between the instants $t=m(t)\tau+s(t)$ and $t=0$. It is the sum of the energy jumps, 
\begin{equation}
\Delta E(t) = \sum_{k=1}^{m(t)} j(k), \mbox{\ \ for $t\geq\tau$,}
\label{m72}
\end{equation}
and $\Delta E(t)=0$ if $0\leq t<\tau$. 
Estimate \fer{m69} shows that for any normal state $\omega$ on $\mm$  there is a constant $C$ s.t. 
\begin{equation}
\left| \frac{\omega(\Delta E(t))}{t}  - \d E_+ \right|\leq \frac{C}{t},
\label{m73}
\end{equation}
for all $t>0$. The energy grows asymptotically linearly in time.

\subsubsection{Entropy, average 2$^{\rm \bf nd}$ law of thermodynamics}
\label{subsectionentropy}

Let $\omega$ and $\omega_0$ be two normal states on $\mm$. The relative entropy of $\omega$ with respect to $\omega_0$ is denoted by ${\rm Ent}(\omega|\omega_0)$, where our definition of relative entropy differs from that one given in \cite{BR} by a sign, so that in our case, ${\rm Ent}(\omega|\omega_0)\geq 0$.

For a thermodynamic interpretation of the entropy and its relation to the energy variation, we assume in this section that $\Omega_\s$ is a $(\beta_\s,\tau_\s^t)$--KMS state on $\mm_\s$, and that $\Omega_\ee$ is a $(\beta_\ee, \tau_\ee^t)$--KMS state on $\mm_\ee$, where $\beta_\s, \beta_\ee$ are inverse temperatures. Let $\omega_0$ be the state on $\mm$ determined by the vector $\Omega_\s\otimes\Omega_\c$ (c.f. before \fer{m1}, and \fer{m16}).

We are interested in the change of relative entropy of the repeated interaction system as time evolves. 
\begin{proposition}
\label{proposition4}
Let $\omega$ be any normal state on $\mm$. Then ${\rm Ent}\big(\omega\circ\alpha_{\rm RI}^t |\omega_0\big)$ is a continuous, piecewise differentiable function of $t\geq 0$. Moreover, we have 
\begin{equation}
{\rm Ent}\big(\omega\circ\alpha_{\rm RI}^t |\omega_0\big) - {\rm Ent}(\omega| \omega_0) =\omega\Big( \beta_\ee \Delta E(t) -\alpha_{\rm RI}^t(X(t)) +X(0)\Big),
\label{m75}
\end{equation}
where $\Delta E(t)$ is the variation of the total energy between the moments $t=0$ and $t=m(t)\tau+s(t)$, \fer{m72}, and where 
\begin{equation}
X(t) = \beta_\ee V_{m(t)+1} + (\beta_\ee-\beta_\s) L_\s,
\label{m75.1}
\end{equation}
with $V_k$ given by \fer{m68.1}.
\end{proposition}
The proof of \fer{m75} is based on the {\it entropy production formula} \cite{JP1}. We give it in Section \ref{sectprop4}. It is not hard to verify that for $t\in (m\tau,(m+1)\tau)$ we have 
\begin{equation}
\frac{\d}{\d t}{\rm Ent}\big(\omega\circ\alpha_{\rm RI}^t |\omega_0\big) =- \omega \Big( \alpha_{\rm RI}^t\big( \i[\beta_\s L_\s +\beta_\ee L_{\ee,m(t)+1}, V_{m(t)+1}]\big)\Big),
\label{m68.2}
\end{equation}
and that left and right derivatives of ${\rm Ent}\big(\omega\circ\alpha_{\rm RI}^t |\omega_0\big)$ exist as $t\rightarrow m\tau$, but they do not coincide.

If ${\rm Ent}(\omega|\omega_0)<\infty$ then all terms in \fer{m75} are bounded uniformly in $t$, except possibly  ${\rm Ent}(\omega\circ\alpha_{\rm RI}^t |\omega_0)$ and $\omega(\beta_\ee\Delta E(t))$. Hence \fer{m82} and \fer{m73} show that for any normal state $\omega$ on $\mm$ there is a constant $C$ s.t. 
\begin{equation}
\left| \frac{{\rm Ent}(\omega\circ\alpha_{\rm RI}^t|\omega_0)}{t} - \frac{\beta_\ee}{\tau} \,\omega_+(j_+)\right| \leq\frac{C}{t},
\label{m76}
\end{equation}
for all $t>0$. The entropy grows linearly in time, for large times.

The relative entropy is non--negative, so \fer{m76} shows that 
\begin{equation}
\omega_+(j_+)\geq 0. 
\label{m77}
\end{equation}
We show in Section \ref{examplesection} that $\omega_+(j_+)$ is strictly positive for concrete systems.  It follows from \fer{m76} also that 
\begin{equation}
\sup_{t\geq 0} \big| {\rm Ent}(\omega\circ\alpha_{\rm RI}^t|\omega_0)\big|<\infty \ \Longleftrightarrow \ \omega_+(j_+)=0.
\label{m78}
\end{equation}
Since $\omega_+(j_+)$ is independent of $\omega$ it follows that for a given interaction ($V,\tau$) the relative entropy either diverges for all initial states $\omega$, as $t\rightarrow\infty$, or it stays bounded for all initial states $\omega$. {\it In particular, if $\omega_+(j_+)>0$ then there does not exist any normal state $\omega$ on $\mm$ which is invariant under $\alpha_{\rm RI}^t$ (i.e., such that $\omega\circ\alpha_{\rm RI}^t=\omega$, for all $t\geq 0$)}.

\begin{proposition}
\label{prop5}
We have 
\begin{equation}
\lim_{t\rightarrow\infty} \big[{\rm Ent}(\omega\circ\alpha_{\rm RI}^{t+\tau}|\omega_0)- {\rm Ent}(\omega\circ\alpha_{\rm RI}^t|\omega_0)\big] = \beta_\ee\,\omega_+(j_+).
\end{equation}
\end{proposition}

The change of entropy during an interval of duration $\tau$, for $t\rightarrow\infty$, is thus given by $\beta_\ee\, \omega_+(j_+)\geq 0$. We call 
\begin{equation}
\d S_+= \frac{\beta_\ee}{\tau}\omega_+(j_+)
\label{m81}
\end{equation}
the (average) {\it asymptotic entropy production}. The quantity $\d S_+$ represents the increase in entropy per unit time, in the limit of large times. It does not depend on the initial state of the system. 

{\it Remark.\ } One sees easily that the expectation of $\d S_+$ is constant in the state $\omega_+\otimes\omega_\c$ (see also \fer{m71}).

Relations \fer{m81} and \fer{m82} lead us to the {\it average {\rm 2$^{\rm nd}$} law of thermodynamics}, 
\begin{equation}
\d E_+ = T_\ee\d S_+,\ \ \ T_\ee=1/\beta_\ee.
\label{m83}
\end{equation}
This law does not depend on the initial state of the system.

%\subsection{Remarks on the assumption $\dim\h_\s<\infty$}

\section{Examples}
\label{examplesection}

\subsection{Spin-Fermion system with quadratic interaction}\label{section:sfq}

As our main example, we consider the case where the small system $\s$
is a $2-$level system and the elements of the chain consist of free
Fermi reservoirs at positive temperature $\beta^{-1}$. Let us first
describe precisely the model (see also \cite{JP} and references therein).

The von Neumann algebra of observables for the small system is 
\beq\label{systalg1}
\mm_\s=M_2(\cc)\otimes \one=\{A\otimes \one | A\in M_2(\cc)\}
\eeq 
acting on the Hilbert space 
\beq\label{systspace1}
\h_\s=\cc^2\otimes \cc^2. 
\eeq 
Let $\sigma_x,$ $\sigma_y,$ $\sigma_z$ be the usual Pauli matrices,
i.e. $\sigma_x=\left(\begin{array}{cc} 0 & 1 \\ 1 & 0
  \end{array}\right),$ $\sigma_y=\left(\begin{array}{cc} 0 & \i \\ -\i &
0 \end{array}\right)$, $\sigma_z=\left(\begin{array}{cc} 1 & 0 \\ 0 &
    -1 \end{array}\right)$. The dynamics of the small system is then given by 
\beq\label{systdyn1}
\tau_\s^t(A\otimes\one)= \e^{\i t\sigma_z}A\e^{-\i t\sigma_z}\otimes\one.
\eeq 
For convenience we chose the reference state $\omega_\s$ to be the
tracial state, i.e. $\omega_\s(A\otimes \one)=\frac{1}{2}\Tr(A)$. Note
that it is a $(\tau_\s^t,0)-$KMS state. Its representative vector is 
\beq\label{systvect1}
\Omega_\s=\frac{1}{\sqrt{2}}\psi_1\otimes\psi_1+\frac{1}{\sqrt{2}}\psi_2\otimes\psi_2
\eeq
where $(\psi_1,\psi_2)$ is the canonical basis of $\cc^2$. For shortness, we will denote by $\psi_{ij}:=\psi_i\otimes\psi_j$ the corresponding basis of $\h_\s$. The standard Liouvillean then writes 
\beq\label{systliouv1} 
L_\s=\sigma_z\otimes\one-\one\otimes\sigma_z,
\eeq
and its spectrum is ${\rm spec}(L_\s)=\{-2,0,2\}$ where $0$ has
multiplicity $2$, and $-2$, $2$ are non degenerate. 
Finally, the modular conjugation and modular operator associated to $(\mm_\s,\Omega_\s)$ are  
\beq\label{systmod1}
J_\s (\phi\otimes\psi) =\bpsi\otimes\bar{\phi}, \quad \Delta_\s=\one\otimes\one, 
\eeq
and where $\bar{\cdot}$ denotes the usual complex conjugation on $\cc^2.$

We then describe an element of the chain, i.e. a free Fermi gas at inverse temperature $\beta$. Let $\fh$ be the Hilbert space of one single fermion and $h$ its energy operator. The operators $a(f)$ and $a^*(f)$ denote the corresponding annihilation and creation operators acting on the fermionic Fock space $\Gamma_-(\fh)$ and they satisfy the canonical anti-commutation relations (CAR). As a consequence of the CAR, the operators $a(f)$ and $a^*(f)$ are bounded and satisfy $\|a^\#(f)\|=\|f\|$ where $a^\#$ stands either for $a$ or for $a^*$. The algebra of observables of a free Fermi gas is the $C^*$-algebra of operators $\fA$ generated by $\{a^\#(f)|f\in\fh\}$. The dynamics is then given by 
\beq
\tau_\f^t(a^\#(f))=a^\#(\e^{\i th}f).
\eeq
It is well known (see e.g. \cite{BR,P}) that for any $\beta>0$, there is a unique $(\tau_\f,\beta)-$KMS state $\omega_\beta$ on $\fA$ which is determined by the two point function $\omega_\beta(a^*(f)a(f))=\langle f, (1+\e^{\beta h})^{-1}f\rangle.$ Finally, let $\Omega_\f$ be the Fock vacuum and $N$ the number operator.

We now fix a complex conjugation (anti-unitary involution) $f\to\bar{f}$ on $\fh$ which commutes with the energy operator $h$. It naturally extends to a complex conjugation on the Fock space $\Gamma_-(\fh)$ and we denote it by the same symbol, i.e. $\Phi\to\bar{\Phi}.$

The GNS representation of the algebra $\fA$ associated to the KMS-state $\omega_\beta$ is the triple $(\h_\F,\pi_\beta,\Omega_\F)$ \cite{AW} where 
\beq
\h_\F=\Gamma_-(\fh)\otimes\Gamma_-(\fh), \quad
\Omega_\F=\Omega_\f\otimes\Omega_\f,
\eeq
and
\beq
\begin{array}{l}
\pi_\beta(a(f)) = a\left(\frac{\e^{\beta h/2}}{\sqrt{1+\e^{\beta h}}}f \right)\otimes \one+(-1)^N\otimes a^*\left(\frac{1}{\sqrt{1+\e^{\beta h}}}\bar{f} \right)=:a_\beta(f),\\
\pi_\beta(a^*(f)) = a^*\left(\frac{\e^{\beta h/2}}{\sqrt{1+\e^{\beta h}}}f \right)\otimes \one+(-1)^N\otimes a\left(\frac{1}{\sqrt{1+\e^{\beta h}}}\bar{f} \right)=:a^*_\beta(f).
\end{array}
\eeq

The von Neumann algebra of observables for an element of the chain will then be the enveloping von Neumann algebra
\beq\label{chainalg1}
\mm_\ee=\pi_\beta(\fA)'',
\eeq
acting on the Hilbert space
\beq\label{chainspace1}
\h_\ee=\h_\F.
\eeq
The dynamics on $\pi_\beta(\fA)$ is given by 
\beq\label{chaindyn1}
\tau_\ee^t(\pi_\beta(A))=\pi_\beta(\tau_\f^t(A))
\eeq 
and extends to $\mm_\ee$ in a unique way. The representative vector of the equilibrium state is
\beq\label{chainvect1}
\Omega_\ee=\Omega_\F,
\eeq
and the standard Liouvillean then writes
\beq\label{chainliouv1}
L_\ee=\d\Gamma(h)\otimes\one-\one\otimes\d\Gamma(h).
\eeq
Finally the modular conjugation and the modular operator associated to $(\mm_\ee,\Omega_\ee)$ are
\beq\label{chainmod1}
J_\ee (\Phi\otimes\Psi)=(-1)^{N(N-1)/2}\bar{\Psi}\otimes (-1)^{N(N-1)/2}\bar{\Phi}, \quad \Delta_\ee=\e^{-\beta L_\ee}.
\eeq

We finally specify the interaction between the small system and the elements of the chain, i.e. the operator $V$. Let $g\in\fh$ be a form factor, we set 
\beq\label{interaction1}
V:=\sigma_x\otimes\one_{\cc^2}\otimes a^*_\beta(g)a_\beta(g)\,\in\mm_\s\otimes\mm_\ee .
\eeq
This is the simplest non trivial interaction for which the number of particles is conserved.

We moreover assume that 
\begin{itemize}
\item[{\bf (SF1)}] $\quad \e^{\beta h/2}g\in\fh$. 
\end{itemize}

\noindent This ensures that Assumption (A) is satisfied. Indeed, using \fer{systmod1}-\fer{chainmod1}-\fer{interaction1}, we get
\begin{eqnarray}\label{cliouvint1}
 & & \Delta^{1/2}V\Delta^{-1/2}\\
 & = & \sigma_x\otimes\one_{\C^2}\otimes\left[ \left( a^*\left(\frac{1}{\sqrt{1+\e^{\beta h}}}g\right)\otimes\one+(-1)^N\otimes a\left(\frac{\e^{-\beta h/2}}{\sqrt{1+\e^{\beta h}}}\bar{g} \right)\right)\right.\nonumber \\
 & & \qquad \qquad \qquad \times\left.\left( a\left(\frac{\e^{\beta h}}{\sqrt{1+\e^{\beta h}}}g\right)\otimes\one+(-1)^N\otimes a^*\left(\frac{\e^{\beta h/2}}{\sqrt{1+\e^{\beta h}}}\bar{g} \right)\right) \right]. \nonumber
\end{eqnarray}
The Liouville operator which generates the interacting dynamics is then the selfadjoint operator
\beq\label{liouv1}
L_\lambda:=L_\s+L_\ee+\lambda V,
\eeq
while the $C$--Liouville operator is 
\beq\label{cliouv1}
K_\lambda:=L_\s+L_\ee+\lambda(V-J\Delta^{1/2}V\Delta^{-1/2}J)=K_0+\lambda W,
\eeq
and where $\lambda\in\R$ is a coupling constant. 

For simplicity reasons we will moreover assume that 
\begin{itemize}
\item[{\bf (SF2) }]  $\fh=L^2(\R^+,{\frak g})$ where ${\frak g}$ is some auxiliary Hilbert space and the operator $h$ is the multiplication operator by $r\in\R^+.$ 
\end{itemize}
Finally, let $g_\beta(r):=(1+\e^{-\beta r})^{-1/2}g(r).$

Note also that the system $\ee$ has the property of return to equilibrium.

Our first result is the 
%%%%%%%%%%%%%%%%%%%%%%% THM: ergodicity for spin-fermion quadratic %%%%%%%%%%%%%%%%%%%%%%%%%%%%%%%%%%%%%%%%%%%%%%%%
\bet\label{thm:ergodic1} Suppose Assumption {\bf (SF1)}-{\bf (SF2)}
are satisfied. Then for any $\tau\notin \frac{\pi}{2}+\pi\N$, there
exists $\Lambda_0 >0$ such that for all $0<|\lambda| < \Lambda_0$, the
operator $M_\lambda:=P\e^{\i \tau K_\lambda}P$ satisfies the ergodic
assumption (E). In particular the spin-fermion system with quadratic
interaction satisfies Theorem \ref{thm1}, with
$\gamma=\frac{\tau^2(\alpha_1+\alpha_2)}{2}\lambda^2+O(\lambda^3)$, Theorem
\ref{thm2}, and moreover the asymptotic state $\omega_{+,\lambda}$ is given by
\begin{eqnarray}\label{asympstate1}
\omega_{+,\lambda}(A_\s) & = & \frac{1}{\alpha_1+\alpha_2}\langle \alpha_1\psi_{11}+\alpha_2\psi_{22},A_\s(\psi_{11}+\psi_{22})\rangle\\
 & & \, +\lambda \left\|\e^{-\beta h/2}g_\beta\right\|_\fh^2\frac{\alpha_1-\alpha_2}{2(\alpha_1+\alpha_2)}\langle \psi_{12}+\psi_{21},A_\s(\psi_{11}+\psi_{12})\rangle+O(\lambda^2),\nonumber
\end{eqnarray}
where, for $j=1,2,$
\beq\label{alphaj1}
\alpha_j:= \int\!\int\d r_1\d r_2\, \e^{-\beta r_j}\|g_\beta(r_1)\|_\fg^2\|g_\beta(r_2)\|_\fg^2\,\sinc^2\left(\frac{\tau(2-r_1+r_2)}{2}\right)>0,
\eeq
$\sinc(x):=\frac{\sin x}{x},$ and all the integrals run over $\R^+$.
\eet
%%%%%%%%%%%%%%%%%%%%%%%%%%%%%%%%%%%%%%%%%%%%%%%%%%%%%%%%%%%%%%%%%%%%%%%%%%%%%%%%%%%%%%%%%%%%%%%%%%%%%%%%%%%%%%%%%%%

The unperturbed operator $M_0$ has eigenvalues $1$ (with multiplicity
$2$), $\e^{2\i\tau}$ and $\e^{-2\i\tau}$ (see \fer{systliouv1}). The assumption on the interaction time $\tau$ ensures that these eigenvalues do not coincide and makes the computation in perturbation theory as simple as possible. However, it can probably be omitted.

One can also see that the asymptotic expectation $E_+(t)$ (see
\fer{m-2}) has a non trivial periodicity (i.e. it is not constant) at
the order $\lambda$.

We now turn to the question of entropy production for this simple model. We will prove that it is strictly positive, at least for small coupling constant. More precisely, we have
%%%%%%%%%%%%%%%%%%%%%%% THM: entropy  for spin-fermion quadratic %%%%%%%%%%%%%%%%%%%%%%%%%%%%%%%%%%%%%%%%%%%%%%%%
\bet\label{thm:entropy1} Suppose Assumptions {\bf (SF1)}-{\bf (SF2)} are satisfied. Then for any $\tau\notin \frac{\pi}{2}+\pi\N$, there exists $\Lambda_1 >0$ such that, for all $0<|\lambda| < \Lambda_1$, the system has strictly positive asymptotic entropy production.
\eet
%%%%%%%%%%%%%%%%%%%%%%%%%%%%%%%%%%%%%%%%%%%%%%%%%%%%%%%%%%%%%%%%%%%%%%%%%%%%%%%%%%%%%%%%%%%%%%%%%%%%%%%%%%%%%%%%%%%

%%%%%%%%%%%%%%%%%%%%%%%%%%%%%%%%%%%%%%%%%%%%%%%%%%%%%%%%%%%%%%%%%%%%%%%%%%%%%%%%%%%%%%%%%%%%%%%%%%%%%%%
%%%%%%%%%%%%%%%%%%%%%%%%%%%%%%%%%%%%%%%%%%%%%%%%%%%%%%%%%%%%%%%%%%%%%%%%%%%%%%%%%%%%%%%%%%%%%%%%%%%%%%%

\subsection{Spin-Fermion system with linear interaction}\label{section:sfl}

As a second example, we consider the same spin fermion system at inverse temperature $\beta$. The only change concerns the interaction term. 
Let $g\in\fh$ be a form factor, we set
\beq\label{interaction2}
V:=\sigma_x\otimes\one_{\cc^2}\otimes(a_\beta(g)+a^*_\beta(g)).
\eeq
Once again, we assume that Assumption (SF1) holds which ensures that $\Delta^{1/2}V\Delta^{-1/2}\in\mm.$ Indeed 
\begin{eqnarray*}
\Delta^{1/2}V\Delta^{-1/2} & = & \sigma_x\otimes\one_{\C^2}\otimes\left[ a^*\left(\frac{1}{\sqrt{1+\e^{\beta h}}}g\right)\otimes\one+a\left(\frac{\e^{\beta h}}{\sqrt{1+\e^{\beta h}}}g\right)\otimes \one \right.\\
 & & \quad \left. +(-1)^N\otimes a^*\left(\frac{\e^{\beta h/2}}{\sqrt{1+\e^{\beta h}}}\bar{g}\right)+(-1)^N\otimes a\left(\frac{\e^{-\beta h/2}}{\sqrt{1+\e^{\beta h}}}\bar{g}\right)\right].
\end{eqnarray*}

\noindent We then have the same kind of result as for the case of quadratic interaction, namely

%%%%%%%%%%%%%%%%%%%%% THM: spin-fermion linear %%%%%%%%%%%%%%%%%%%%%%%%%%%%%%%%%%%%%%%%%%%%%%%%%%%%%%%%
\bet\label{thm:ergodic2} Suppose Assumption {\bf (SF1)}-{\bf (SF2)}
are satisfied. Then for any $\tau\notin \frac{\pi}{2}+\pi\N$, there
exists $\Lambda_0 >0$ such that for all $0<|\lambda| < \Lambda_0$, the
operator $M_\lambda:=P\e^{\i \tau K_\lambda}P$ satisfies the ergodic
assumption (E). In particular the spin-fermion system with linear
interaction satisfies Theorem \ref{thm1}, with $\gamma=\frac{\tau^2(\alpha_1+\alpha_2)}{2}\lambda^2+O(\lambda^3)$, Theorem \ref{thm2} and the asymptotic state $\omega_{+,\lambda}$ is given by
\beq\label{asympstate2}
\omega_{+,\lambda}(A_\s)=\frac{1}{\alpha_1+\alpha_2}\langle \alpha_1\psi_{11}+\alpha_2\psi_{22},A_\s(\psi_{11}+\psi_{22})\rangle+O(\lambda^2),
\eeq
where
\begin{eqnarray*}\label{alphaj2}
\alpha_1 & := & \int\d r\, \|g_\beta(r)\|_\fg^2\, \left(\e^{-\beta r}\sinc^2\left(\frac{\tau(r-2)}{2}\right)+\sinc^2\left(\frac{\tau(r+2)}{2}\right)\right),\\
\alpha_2 & := & \int\d r\, \|g_\beta(r)\|_\fg^2\, \left(\e^{-\beta r}\sinc^2\left(\frac{\tau(r+2)}{2}\right)+\sinc^2\left(\frac{\tau(r-2)}{2}\right)\right).
\end{eqnarray*}
and all the integrals run over $\R^+$. Moreover the system has
strictly positive asymptotic entropy production.
\eet
%%%%%%%%%%%%%%%%%%%%%%%%%%%%%%%%%%%%%%%%%%%%%%%%%%%%%%%%%%%%%%%%%%%%%%%%%%%%%%%%%%%%%%%%%%%%%%%%%%%%%%%

%%%%%%%%%%%%%%%%%%%%%%%%%%%%%%%%%%%%%%%%%%%%%%%%%%%%%%%%%%%%%%%%%%%%%%%%%%%%%%%%%%%%%%%%%%%%%%%%%%%%%%%%%%%%%%%%%
%%%%%%%%%%%%%%%%%%%%%%%%%%%%%%%%%%%%%%%%%%%%%%%%%%%%%%%%%%%%%%%%%%%%%%%%%%%%%%%%%%%%%%%%%%%%%%%%%%%%%%%%%%%%%%%%%

\subsection{Spin-Spin model}\label{section:ss}

As our last example, we consider a model in which the small system as well as the elements of the chain consist of a $2-$level system. Such kind of systems (or more generally a $d-$level system interacting with a chain of $n-$level systems) have been considered previously in \cite{AJ}.

The von Neumann algebra of observables for the small system and for the elements of the chain is 
\beq\label{systalg3}
\mm_\s=\mm_\ee=M_2(\cc)\otimes \one=\{A\otimes \one | A\in M_2(\cc)\}
\eeq 
acting on the Hilbert space 
\beq\label{systspace3}
\h_\s=\h_\ee=\cc^2\otimes \cc^2. 
\eeq 
Let $E_\s$ and $E_\ee$ be non negative real numbers. They will play the role of the energy of the ``excited'' state of the small system and of the elements of the chain respectively. 
The dynamics of the small system is then given by 
\beq\label{systdyn3}
\tau_\s^t(A\otimes\one)= \e^{\i t h_\s}A\e^{-\i t h_\s}\otimes\one,
\eeq 
and the one of an element of the chain by 
\beq\label{chaindyn3}
\tau_\ee^t(A\otimes\one)= \e^{\i t h_\ee}A\e^{-\i t h_\ee}\otimes\one,
\eeq
where $h_\s=\left( \begin{array}{cc} 0 & 0 \\ 0 & E_\s \end{array}\right)$ and $h_\ee=\left( \begin{array}{cc} 0 & 0 \\ 0 & E_\ee \end{array}\right).$

Once again, for convenience, we chose the reference state $\omega_\s$ to be the tracial state, i.e. $\omega_\s(A\otimes \one)=\frac{1}{2}\Tr(A)$ (the results of course do not depend on this choice). Its representative vector is (Section \ref{section:sfq})
\beq\label{systvect3}
\Omega_\s=\frac{1}{\sqrt{2}}\psi_{11}+\frac{1}{\sqrt{2}}\psi_{22}.
\eeq
The standard Liouvillean writes 
\beq\label{systliouv3} 
L_\s=h_\s \otimes\one-\one\otimes h_\s,
\eeq
and the modular conjugation and modular operator associated to $(\mm_\s,\Omega_\s)$ are  
\beq\label{systmod3}
J_\s (\phi\otimes\psi) =\bpsi\otimes\bar{\phi}, \quad \Delta_\s=\one\otimes\one. 
\eeq

In order to avoid confusions between the small system and an element of the chain we will denote by $\phi_{ij}=\phi_i\otimes\phi_j$ instead of $\psi_{ij}$ the basis of $\h_\ee$.
The reference state $\omega_\ee$ will be the $(\tau_\ee,\beta)-$KMS state. Its representative vector writes 
\beq\label{chainvect3}
\Omega_\ee=\frac{1}{\sqrt{1+\e^{-\beta E_\ee}}}(\phi_{11}+\e^{-\beta E_\ee/2}\phi_{22}).
\eeq
The standard Liouville operator is 
\beq\label{chainliouv3} 
L_\ee=h_\ee \otimes\one-\one\otimes h_\ee,
\eeq
and the modular conjugation and modular operator associated to $(\mm_\ee,\Omega_\ee)$ are  
\beq\label{chainmod3}
J_\ee (\phi\otimes\psi) =\bpsi\otimes\bar{\phi}, \quad \Delta_\ee=\e^{-\beta L_\ee}. 
\eeq
Note that here the system $\ee$ does \emph{not} have the property of return to equilibrium.

We now describe the interaction. Let us denote by $a$ and $a^*$ the annihilation and creation operators associated to the vectors $\phi_1$ (ground state) and $\phi_2$ (excited state), i.e. $a\phi_1=0, a\phi_2=\phi_1, a^*\phi_1=\phi_2, a^*\phi_2=0.$ Finally let $I=\left(\begin{array}{cc} \eba & \ebb \\ \ebc & \ebd \end{array}\right)\in M_2(\cc).$ The interaction is then given by 
\beq\label{interaction3}
V=I\otimes \one \otimes a^*\otimes \one+I^*\otimes \one \otimes a\otimes \one.
\eeq
The Liouville operator which generates the interacting dynamics is then the selfadjoint operator
\beq\label{liouv3}
L_\lambda:=L_\s+L_\ee+\lambda V,
\eeq
while the $C$--Liouville operator is 
\beq\label{cliouv3}
K_\lambda:=L_\s+L_\ee+\lambda(V-J\Delta^{1/2}V\Delta^{-1/2}J)=K_0+\lambda W,
\eeq
where $\lambda$ is a coupling constant. 

We finally consider the following assumptions. 
\begin{itemize}
\item[{\bf (SS1)}] $\ebb\neq 0$ and $\tau(E_\ee-E_\s)\notin 2\pi\Z.$ 
\end{itemize}
\begin{itemize}
\item[{\bf (SS2)}] $\ebc\neq 0$ and $\tau(E_\ee+E_\s)\notin 2\pi\Z.$ 
\end{itemize}
If either {\bf (SS1)} or {\bf (SS2)} is satisfied, then the ergodic assumption (E) holds.

%%%%%%%%%%%%%%%%%%%%% THM: spin-spin %%%%%%%%%%%%%%%%%%%%%%%%%%%%%%%%%%%%%%%%%%%%%%%%%%%%%%%%%%%%%%%%%%%%%
\bet\label{thm:ergodic3} Suppose that $\tau E_\ee \notin \pi\Z$ and
that either Assumption {\bf (SS1)} or {\bf (SS2)} is satisfied. Then,
there exists $\Lambda_0 >0$ such that for all $0<|\lambda| <
\Lambda_0$, the operator $M_\lambda:=P\e^{\i \tau K_\lambda}P$
satisfies the ergodic assumption (E). In particular the spin-spin
system satisfy Theorem \ref{thm1}, with
$\gamma=\gamma_0\lambda^2+O(\lambda^3)$, and the asymptotic state $\omega_{+,\lambda}$ is given by
\beq\label{asympstate3}
\omega_{+,\lambda}(A_\s)=\frac{1}{\alpha_1+\alpha_2}\langle \alpha_1\psi_{11}+\alpha_2\psi_{22},A_\s(\psi_{11}+\psi_{22})\rangle+O(\lambda^2),
\eeq
where
\begin{eqnarray*}\label{alphaj3}
\alpha_1 & := & |\ebb|^2 \sinc^2 \left( \frac{\tau(E_\ee-E_\s)}{2} \right)+\e^{-\beta E_\ee} |\ebc|^2 \sinc^2 \left( \frac{\tau(E_\ee+E_\s)}{2}\right)\geq 0,\\
\alpha_2 & := & \e^{-\beta
  E_\ee}|\ebb|^2\sinc^2\left(\frac{\tau(E_\ee-E_\s)}{2}\right)+|\ebc|^2\sinc^2\left(\frac{\tau(E_\ee+E_\s)}{2}\right)\geq 0,\\
\gamma_0 & := & \min
\left(\frac{\tau^2(\alpha_1+\alpha_2)}{1+\e^{-\beta E_\ee}},
  \frac{\tau^2(\alpha_1+\alpha_2)}{2(1+\e^{-\beta E_\ee})}\right.\\
 & & \qquad\qquad\qquad\qquad\qquad   \left. +\frac{\tau^2}{2}\sinc^2(\frac{\tau E_\ee}{2})(|\eba|^2+|\ebd|^2-\bar{\eba}\ebd-\eba\bar{\ebd}) \right).
\end{eqnarray*}
If moreover both Assumptions {\bf (SS1)} and {\bf (SS2)} are
satisfied, then the system has strictly positive asymptotic entropy production.
\eet
%%%%%%%%%%%%%%%%%%%%%%%%%%%%%%%%%%%%%%%%%%%%%%%%%%%%%%%%%%%%%%%%%%%%%%%%%%%%%%%%%%%%%%%%%%%%%%%%%%%%%%%

Once again, we assume that $\tau E_\ee\notin \pi\Z$ just in order to make the eigenvalues of $M_0$ not coincide and this can probably be weakened. On the other hand, Assumptions {\bf (SS1)} and {\bf (SS2)} are much deeper. Their signification is that there is an effective coupling between the ground state and the excited state of the small system ($\ebb$ or $\ebc$ non zero) as well as a non resonant phenomenon between the energies of the small system and the elements of the chain ($\tau(E_\ee-E_\s)$ or $\tau(E_\ee+E_\s)$ not in $2\pi\Z$). Asking that either {\bf (SS1)} or {\bf (SS2)} is satisfied is actually equivalent to the condition $\alpha_1+\alpha_2\neq 0.$

\section{Proofs}

\subsection{Proofs of Theorems \ref{thm1}, \ref{thm2}}
\label{msect2}
We give the full proof of Theorem \ref{thm2}, the proof of Theorem \ref{thm1} is a special case of the former.

It is enough to show \fer{m-3.0}, \fer{m-3} for vector states $\omega(\cdot)=\scalprod{\psi}{\cdot \ \psi}$, $\psi\in\h$, $\|\psi\|=1$. Further, since every $\psi\in\h$ is approximated in the norm of $\h$ by finite linear combinations of vectors of the form $\psi_\s\otimes_{m\geq 1}\psi_m$, where $\psi_\s\in\h_\s$, $\psi_m=\Omega_\ee$ if $m>N$, for some $N<\infty$, it suffices to prove \fer{m-3.0}, \fer{m-3} for vector states determined by vectors of the form
\begin{equation}
\psi_\s\otimes_{m=1}^N\psi_m\otimes_{m>N}\Omega_\ee\in\h,
\label{m24}
\end{equation}
where $\|\psi_\s\|=\|\psi_m\|=1$, $1\leq m\leq N$, for arbitrary  $N<\infty$. Finally, since the vectors $\Omega_\s$, $\Omega_\ee$ are cyclic for the commutants $\mm_\s'$, $\mm_\ee'$, any vector of the form \fer{m24} is approximated by a
\begin{equation}
\psi = B'\ \Omega_\s\otimes\Omega_\c,
\label{m25}
\end{equation}
for some 
\begin{equation}
B'= B_\s'\otimes_{m=1}^N B_m'\otimes_{m>N}\bbbone_\ee\ \in\mm',
\label{m26}
\end{equation}
with $B_\s'\in\mm_\s'$, $B_m'\in\mm_\ee'$. It is therefore sufficient to show \fer{m-3.0}, \fer{m-3} for vectors of the form \fer{m25}, \fer{m26}. 

\bigskip
Let $A_\s\in\mm_\s$, $A_1,\ldots, A_p\in\mm_\ee$ and $B_{-\ell},\ldots,B_0,\ldots, B_r \in\mm_\ee$ be fixed observables ($\ell, r\geq 0$). We examine the expectation value
\begin{equation}
E(t):= \scalprod{\psi}{\alpha_{\rm RI}^t\left( A_\s\otimes_{i=1}^p A_i\otimes_{j=-\ell}^{r}\vartheta_{m(t)+j+1}(B_j)\right)\psi},
\label{m27}
\end{equation}
where $\psi$ is given in by \fer{m25}, $m(t)$ is determined by \fer{m9}, and where the $A_i$ act on the first $p$ factors of $\h_\c$. It is clear that the trivial factors $\bbbone_\ee$ are omitted in \fer{m27}. The choice of indices in \fer{m27} is such that at time $t$ the observable $B_0$ is measured in the element $\ee$ of the chain which is interacting with $\s$ (i.e., the $(m(t)+1)$--th element of the chain). The following decomposition serves to isolate the dynamics of the elements $\ee$ which do not interact at a given time, see also \fer{m8}.
\begin{equation}
\e^{-\i s\tilde{L}_{m+1}}\e^{-\i\tau\tilde{L}_m}\cdots\e^{-\i\tau \tilde{L}_1}=  U^-_m \e^{-\i s{L}_{m+1}}\e^{-\i\tau{L}_m}\cdots\e^{-\i\tau {L}_1}U^+_m,
\label{m28}
\end{equation}
where
\begin{eqnarray}
U^-_m &=& \exp\left[-\i \sum_{j=1}^m[(m-j)\tau+s]L_{\ee,j}\right],
\label{m29}\\
U^+_m &=& \exp\left[-\i \sum_{j=2}^{m+1}(j-1)\tau L_{\ee,j}-\i(m\tau+s) \sum_{j\geq m+2} L_{\ee,j}\right].
\label{m30}
\end{eqnarray}
We obtain from \fer{m10}, \fer{m11} and \fer{m28}
\begin{eqnarray}
\lefteqn{E(t)= \left\langle \psi, (U_m^+)^*\e^{\i\tau L_1}\cdots\e^{\i\tau L_m}\e^{\i s L_{m+1}}\right.}\nonumber\\
&&\times A_\s\otimes_{i=1}^p \tau_\ee^{(m-i)\tau+s}(A_i)\otimes_{j=-\ell}^{-1}\vartheta_{m+j+1}(\tau_\ee^{(-j-1)\tau+s}(B_j))\otimes_{j=0}^r\vartheta_{m+j+1}(B_j) \nonumber\\
&&\left.\times \e^{-\i sL_{m+1}}\e^{-\i\tau L_m}\cdots \e^{-\i\tau L_1}U_m^+\psi\right\rangle,
\label{m31}
\end{eqnarray}
where we write $m, s$ for $m(t), s(t)$. The operator $(U_m^+)^*\cdots U_m^+$ in the r.h.s. of \fer{m31} belongs to $\mm$. Hence we can commute the $B'$ in \fer{m25} with this operator to obtain
\begin{eqnarray}
\lefteqn{E(t)= \left\langle \Omega, (B')^*B' (U_m^+)^*\e^{\i\tau L_1}\cdots\e^{\i\tau L_m}\e^{\i s L_{m+1}}\right.}\nonumber\\
&&\times A_\s\otimes_{i=1}^p \tau_\ee^{(m-i)\tau+s}(A_i)\otimes_{j=-\ell}^{-1}\vartheta_{m+j+1}(\tau_\ee^{(-j-1)\tau+s}(B_j))\otimes\vartheta_{m+1}(B_0) \nonumber\\
&&\left.\times \e^{-\i sL_{m+1}}\e^{-\i\tau L_m}\cdots \e^{-\i\tau L_1}\Omega\right\rangle \prod_{j=1}^r\av{B_j}_{\Omega_\ee},
\label{m32}
\end{eqnarray}
where we have set
\begin{equation}
\Omega=\Omega_\s\otimes\Omega_\c,
\label{m33}
\end{equation}
and we write $\av{O}_{\chi} =\scalprod{\chi}{O \chi}$ for an operator $O$ and a vector $\chi$. The operator $U_m^+$ disappears because it leaves $\Omega$ invariant, c.f. \fer{m7}. We are able to factorize the averages $\av{B_j}_{\Omega_\ee}$, for $j\geq 1$, because $B'$ and all propagators in \fer{m31} act trivially on factors of $\h_\c$ with index $\geq m+2$ (note also that $N<m$ since we have in mind the limit $m\rightarrow\infty$). 

Since $(B')^*B'$ acts trivially on factors in $\h_\c$ with index $\geq N+1$ we may replace $(U_m^+)^*=(U_{m(t)}^+)^*$ in \fer{m32} by 
\begin{equation}
\tilde{U}_N = \exp\left[ \i\sum_{j=1}^N(j-1)\tau L_{\ee,j}\right],
\label{m34}
\end{equation}
which is a unitary {\it not depending} on $t$. Without changing the value of \fer{m32} we can replace in that equation successively $L_1$ by $K_1$, then $L_2$ by $K_2$, up to replacing $L_p$ by $K_p$. Then
\begin{eqnarray}
\lefteqn{E(t)= \big\langle \Omega, (B')^*B' \tilde{U}_N \,\e^{\i\tau K_1}\cdots\e^{\i\tau K_p}\e^{\i\tau L_{p+1}} \cdots\e^{\i\tau L_m}\e^{\i s L_{m+1}}}\nonumber\\
&&\times A_\s\otimes_{j=-\ell}^{-1}\vartheta_{m+j+1}(\tau_\ee^{(-j-1)\tau+s}(B_j))\otimes\vartheta_{m+1}(B_0) \nonumber\\
&&\left.\times \e^{-\i sL_{m+1}}\e^{-\i\tau L_m}\cdots \e^{-\i\tau L_{p+1}}\big(\otimes_{i=1}^p \e^{\i[(m-i)\tau+s]L_\ee}A_i\big)\Omega\right\rangle\prod_{j=1}^r\av{B_j}_{\Omega_\ee},\ \ \ 
\label{m35}
\end{eqnarray}
where we have used that $\Omega$ is in the kernel of the $K_j$, and where we have commuted the product of the freely evolved $A_j$'s to the right through the propagators which act on different factors. 

Because the system $\ee$ has the property of return to equilibrium the propagator $\e^{\i [(m(t)-i)\tau +s(t)]L_\ee}$ converges to the projection $P_{\Omega_\ee}=|\Omega_\ee\rangle\langle\Omega_\ee|$, as $t\rightarrow\infty$, in the weak sense on $\h_\ee$ . Define the orthogonal projection  $Q_p^\perp=\bbbone-Q_p$ on $\h$, where $Q_p=\otimes_{j=1}^p \vartheta_j(P_{\Omega_\ee})$. We want to show that 
\begin{eqnarray}
\lefteqn{\lim_{t\rightarrow\infty} \big\langle \Omega, (B')^*B' \tilde{U}_N \,\e^{\i\tau K_1}\cdots\e^{\i\tau K_p}\e^{\i\tau L_{p+1}} \cdots\e^{\i\tau L_m}\e^{\i s L_{m+1}}}\nonumber\\
&&\times A_\s\otimes_{j=-\ell}^{-1}\vartheta_{m+j+1}(\tau_\ee^{(-j-1)\tau+s}(B_j))\otimes\vartheta_{m+1}(B_0) \nonumber\\
&&\left.\times \e^{-\i sL_{m+1}}\e^{-\i\tau L_m}\cdots \e^{-\i\tau L_{p+1}}Q_p^\perp \big(\otimes_{i=1}^p \e^{\i[(m-i)\tau+s]L_\ee}A_i\big)\Omega\right\rangle=0.
\label{m36}
\end{eqnarray}
To do so we split the Hilbert space as $\h=\h_1\otimes\h_2$, where
$
\h_1=\h_\s\otimes_{m\geq p+1}\h_\ee,
$ 
and $\h_2=\otimes_{i=1}^p\h_\ee$, and set
\begin{eqnarray*}
\psi_1^{m}&=&\e^{\i\tau L_{p+1}} \cdots\e^{\i\tau L_m}\e^{\i s L_{m+1}} \big(A_\s\otimes_{j=-\ell}^{-1}\vartheta_{m+j+1}(\tau_\ee^{(-j-1)\tau+s}(B_j))\nonumber\\
&&\otimes\vartheta_{m+1}(B_0)\big) \e^{-\i sL_{m+1}}\e^{-\i\tau L_m}\cdots \e^{-\i\tau L_{p+1}}\big[\Omega_\s\otimes_{m\geq p+1}\Omega_\ee\big] \in \h_1,
\label{m38}\\
\psi_2^{m}&=& Q^\perp_p \big(\otimes_{i=1}^p \e^{\i[(m-i)\tau+s]L_\ee}A_i\big)\otimes_{i=1}^p\Omega_\ee\in \h_2.
\label{m39}
\end{eqnarray*}
$Q_p^\perp$ is a sum of terms, each one containing the operator $P^\perp_{\Omega_\ee}=\bbbone_\ee-P_{\Omega_\ee}$ acting on at least one of the $p$ factors in $\h_2$. Consequently we have $\psi_2^{m(t)}\rightarrow 0$, weakly in $\h_2$, as $t\rightarrow\infty$. Since $\psi_1^{m(t)}$ is uniformly bounded in $t$ it follows that $\psi_1^{m(t)}\otimes\psi_2^{m(t)}$ converges weakly to zero in $\h$, as $t\rightarrow\infty$. This proves relation \fer{m36}.

Thus, in the limit $t\rightarrow\infty$, the only contribution to \fer{m35} comes from the part where all the free propagators $\e^{\i[(m-i)\tau+s]L_\ee}$ are replaced by $P_{\Omega_\ee}$. This shows that 
\begin{eqnarray}
\lefteqn{
\lim_{t\rightarrow\infty}\Big| E(t) -  \big\langle \Omega, (B')^*B' \tilde{U}_N \,\e^{\i\tau K_1}\cdots\e^{\i\tau K_p}\e^{\i\tau L_{p+1}} \cdots\e^{\i\tau L_m}\e^{\i s L_{m+1}}}\nonumber\\
&&\times A_\s\otimes_{j=-\ell}^{-1}\vartheta_{m+j+1}(\tau_\ee^{(-j-1)\tau+s}(B_j))\otimes\vartheta_{m+1}(B_0)\nonumber\\
&&\times \e^{-\i sL_{m+1}}\e^{-\i\tau L_m}\cdots \e^{-\i\tau L_{p+1}}\Omega\big\rangle\prod_{i=1}^p\av{A_i}_{\Omega_\ee}\prod_{j=1}^r\av{B_j}_{\Omega_\ee}\Big|=0. 
\label{m40}
\end{eqnarray}
We may now, as we did above, turn the operators $L_j$ in \fer{m40} into $K_j$'s, also for the remaining indices $j=p+1,\ldots,m+1$, to arrive at
\begin{eqnarray}
\lefteqn{
\lim_{t\rightarrow\infty}\Big| E(t) -  \big\langle \Omega, (B')^*B' \tilde{U}_N \, \e^{\i\tau K_1}\cdots \e^{\i\tau K_N} P_{N} \e^{\i\tau K_{N+1}}\cdots \e^{\i\tau K_m} \e^{\i s K_{m+1}}}\nonumber\\
&& \ \ \ \ \ \ \ \ \ \ \times A_\s\otimes_{j=-\ell}^{-1}\vartheta_{m+j+1}(\tau_\ee^{(-j-1)\tau+s}(B_j))\otimes\vartheta_{m+1}(B_0)\Omega\big\rangle\nonumber\\
&&\ \ \ \ \ \ \ \ \ \ \times\prod_{i=1}^p\av{A_i}_{\Omega_\ee}\prod_{j=1}^r\av{B_j}_{\Omega_\ee}\Big|=0,
\label{m41}
\end{eqnarray}
where we introduce the projection $P_{N}= \otimes_{m\geq N+1}\vartheta_m(P_{\Omega_\ee})$ (that projection comes from the left factor of the inner product and slips through $(B')^*B'\tilde{U}_N$ and through the first $N$ propagators).

We have 
\begin{equation}
\vartheta_{m+1}(P_{\Omega_\ee})\e^{\i s K_{m+1}} \big( A_\s\otimes\vartheta_{m+1}(B_0)\big) \Omega = D_0(s)\Omega,
\label{m43}
\end{equation}
where $D_0(s)$ is a linear operator acting nontrivially only on $\h_\s$. $D_0(s)$ depends on $A_\s$, $B_0$, the interaction $V$ and $s=s(t)$, but it is {\it independent} of $m=m(t)$. 
%Moreover, $D_0(s)\in\mm_\s$, because in addition to the information $D_0(s)\in\b(\h_\s)$, we know that  
%\begin{eqnarray}
%\lefteqn{
%\vartheta_{m+1}(P_{\Omega_\ee})\e^{\i s K_{m+1}} \big( A_\s\otimes\vartheta_{m+1}(B_0)\big) \Omega}\nonumber\\
% &=& \vartheta_{m+1}(P_\ee)\e^{\i s L_{m+1}} \big( A_\s\otimes\vartheta_{m+1}(B_0)\big)\e^{-\i s L_{m+1}} \Omega \nonumber\\
%&\in & \vartheta_{m+1}(P_\ee)\ \mm\ \Omega\nonumber\\
%&\subset & \mm \ \Omega.
%\label{m44}
%\end{eqnarray}
In the same way we define $D_1(s)\in{\cal B}(\h_\s)$ by 
\begin{equation}
\vartheta_m(P_{\Omega_\ee}) \e^{\i\tau K_m} \vartheta_m(\tau_\ee^{s}(B_{-1})) D_0(s)\Omega = D_1(s) D_0(s)\Omega,
\label{m45}
\end{equation}
and then $D_2(s),\ldots, D_\ell(s)\in{\cal B}(\h_\s)$. Hence the inner product in \fer{m41} can be written as 
\begin{equation}
\scalprod{\Omega}{(B')^*B' \tilde{U}_N \, \e^{\i\tau K_1}\cdots \e^{\i\tau K_N} P_{N} \e^{\i\tau K_{N+1}}\cdots \e^{\i\tau K_{m-\ell}}D(s)\Omega},
\label{m46}
\end{equation}
where $D(s)=D_\ell(s)\cdots D_0(s)\in{\cal B}(\h_\s)$. Since $\Omega=P\Omega$, where $P$ is the projection onto $\Omega_\c$, c.f. \fer{P}, we have  
\begin{equation}
P_{N} \e^{\i\tau K_{N+1}}\cdots \e^{\i\tau K_{m-\ell}}D(s)\Omega
=
P \e^{\i\tau K_{N+1}}\cdots \e^{\i\tau K_{m-\ell}}PD(s)\Omega.
\label{m47}
\end{equation}
The reduced product of the propagators on the r.h.s. is the product of the reduced propagators, as we show in the following proposition.
\begin{proposition}
\label{proposition2}
For any $q\geq 1$, let $t_1,\ldots t_q\in\r$, and let $m_1, \ldots, m_q\geq 1$ be distinct integers. Then we have 
\begin{equation}
P\e^{\i t_1 K_{m_1}}\cdots \e^{\i t_q K_{m_q}}P = P\e^{\i t_1 K_{m_1}}P\cdots P\e^{\i t_q K_{m_q}}P.
\label{m48}
\end{equation}
\end{proposition}
{\it Proof of Proposition \ref{proposition2}.\ }
If $Q$ is a projection we set $Q^\perp=\bbbone-Q$. We have  $P^\perp \e^{\i t_q K_{m_q}}P\in \ran\vartheta_{m_q}(P_\ee^\perp)$ since on all factors of $\h_\c$ with label $m\neq m_q$ the projection $\vartheta_m(P_\ee)$ coming from $P$ commutes with $\e^{\i t_q K_{m_q}}$. It follows that $P\e^{\i t_1 K_{m_1}}\cdots P^\perp\e^{\i t_q K_{m_q}}P = P\e^{\i t_1 K_{m_1}}\cdots \vartheta_{m_q}(P_\ee^\perp)P^\perp\e^{\i t_q K_{m_q}}P=0$, because $\vartheta_{m_p}(P^\perp_\ee)$ can be commuted to the left to hit $P$. Thus we have $P\e^{\i t_1 K_{m_1}}\cdots \e^{\i t_q K_{m_q}}P = P\e^{\i t_1 K_{m_1}}\cdots \e^{\i t_{q-1} K_{m_{q-1}}} P\e^{\i t_q K_{m_q}}P$ and we can repeat the argument. This proves the proposition. \hfill $\blacksquare$

According to \fer{m48} the r.h.s. of \fer{m47} equals $M^{m-\ell-N}D(s)\Omega$, where $M=PMP$ is the operator introduced in \fer{m17}. We then obtain the following result from \fer{m47}, \fer{m46} and \fer{m41}:
\begin{eqnarray}
\lefteqn{
\lim_{t\rightarrow\infty}\Big| E(t) -  \scalprod{\Omega}{(B')^*B' \tilde{U}_N \, \e^{\i\tau K_1}\cdots \e^{\i\tau K_N} PM^{m(t)-\ell-N}P D(s(t))\Omega}}\nonumber\\
&&\ \ \ \ \ \ \ \ \ \ \ \ \ \ \ \ \ \ \ \ \ \ \ \ \ \ \ \ \ \ \ \ \ \ \ \ \times \prod_{i=1}^p\av{A_i}_{\Omega_\ee}\prod_{j=1}^r\av{B_j}_{\Omega_\ee}\Big|=0.\ \ \ \ \ \ \ \ \ \ \ \ \ \ 
\label{m49}
\end{eqnarray}
In order to further simplify the scalar product in \fer{m49} we use the ergodicity assumption (E). We have $M^{m(t)-\ell-N}\rightarrow \pi=|\Omega_\s\rangle\langle \Omega_\s^*|$, as $t\rightarrow\infty$, in the topology of $\b(\h_\s)$ (c.f. \fer{m17.1}), and since $D(s(t))$ is uniformly bounded in $t$ we obtain 
\begin{equation}
\lim_{t\rightarrow\infty}\Big| E(t) -  \scalprod{\Omega}{(B')^*B'\Omega}\scalprod{\Omega_\s^*}{P D(s(t))P\Omega_\s}\prod_{i=1}^p\av{A_i}_{\Omega_\ee}\prod_{j=1}^r\av{B_j}_{\Omega_\ee}\Big|=0.
\label{m50}
\end{equation}
Remember that $\psi$ is approximated by $B'\Omega$, i.e., given an arbitrary $\varepsilon>0$ we choose $B'$ s.t. $\|\psi-B'\Omega\|\leq\varepsilon$. Thus 
\begin{equation}
\scalprod{\Omega}{(B')^*B'\Omega}=\|B'\Omega\|^2=(\|\psi\|+O(\varepsilon))^2 = 1+O(\varepsilon),
\label{m51}
\end{equation}
which we can use in \fer{m50} to arrive at 
\begin{equation}
\lim_{t\rightarrow\infty}\Big| E(t) -  \scalprod{\Omega_\s^*}{P D(s(t))P\Omega_\s}\prod_{i=1}^p\av{A_i}_{\Omega_\ee}\prod_{j=1}^r\av{B_j}_{\Omega_\ee}\Big|=0.
\label{m52}
\end{equation}
Note that if $1$ is a degenerate eigenvalue of $M$, \fer{m50} becomes
\beq\label{m52.1}
\lim_{t\rightarrow\infty}\Big| E(t) -  \scalprod{\Omega}{(B')^*B'\pi P D(s(t))P\Omega_\s}\prod_{i=1}^p\av{A_i}_{\Omega_\ee}\prod_{j=1}^r\av{B_j}_{\Omega_\ee}\Big|=0.
\eeq
Thus, $E(t)$ has a $\tau-$periodic asymptotic behaviour ($s(t)$ is
periodic) but which depends a priori on $B'$ and thus on the initial
state $\omega$: the
simplification due to \fer{m51} does not hold anymore.

Finally by the definition of $D(s)$, c.f. \fer{m46}, \fer{m45}, \fer{m40} we get
\begin{eqnarray}
\lefteqn{
\scalprod{\Omega_\s^*}{P D(s(t))P\Omega_\s}}\label{m53}\\
&=& \Big\langle \Omega_\s^*, P\e^{\i\tau K_1}\cdots\e^{\i\tau K_\ell}\e^{\i sK_{\ell+1}}\big[ A_\s\otimes\tau_\ee^{(\ell-1)\tau+s}(B_{-\ell})\otimes\cdots\otimes \tau_\ee^s(B_{-1})\otimes B_0\big]P\Omega_\s\Big\rangle
\nonumber
\end{eqnarray}
Using the invariance $P\Omega_\s=\e^{-\i sK_{\ell+1}}\e^{-\i\tau K_\ell}\cdots \e^{-\i\tau K_1}P\Omega_\s$ and passing from $K$'s to $L$'s we obtain
\begin{eqnarray}
\lefteqn{\scalprod{\Omega_\s^*}{P D(s(t))P\Omega_\s}}\nonumber\\
&=& \Big\langle \Omega_\s^*, P\e^{\i\tau L_1}\cdots\e^{\i\tau L_\ell}\e^{\i sL_{\ell+1}}\big[ A_\s\otimes\tau_\ee^{(\ell-1)\tau+s}(B_{-\ell})\otimes\cdots\nonumber\\
&&\quad \cdots \otimes \tau_\ee^s(B_{-1})\otimes B_0\big]\e^{-\i sL_{\ell+1}}\e^{-\i\tau L_\ell}\cdots \e^{-\i\tau L_1} P\Omega_\s\Big\rangle\nonumber\\
&=&\Big\langle \Omega_\s^*, P\e^{\i\tau L_1}\cdots\e^{\i\tau L_\ell}\e^{\i sL_{\ell+1}} (U_{\ell}^-)^*\big[ A_\s\otimes B_{-\ell}\otimes\cdots\otimes B_0\big] U_{\ell}^- \nonumber\\
&&\quad \times \e^{-\i sL_{\ell+1}}\e^{-\i\tau L_\ell}\cdots \e^{-\i\tau L_1} P\Omega_\s\Big\rangle\nonumber\\
&=&\scalprod{\Omega_\s^*}{ P \alpha_{\rm RI}^{\ell\tau +s(t)}\left(A_\s\otimes B_{-\ell}\otimes\cdots\otimes B_0\right)P\Omega_\s}.
\label{m54}
\end{eqnarray}
We conclude the proof of \fer{m-3.0}, \fer{m-3} and hence the proof of Theorem \ref{thm2} by plugging \fer{m54} into \fer{m52}.\hfill $\blacksquare$

\subsection{Proof of Theorem \ref{thm3}}
The proof is a simple modification of the proof of Theorem \ref{thm2}. We indicate the main steps. Proceeding as in Section \ref{msect2} we see that (compare with \fer{m35})
\begin{equation}\label{m65}
\C(t;A,A_\s,B_0)= \scalprod{\Omega}{(B')^*B'A\, \tilde U_N \e^{\i\tau K_1}\cdots \e^{\i\tau K_m}\e^{\i sK_{m+1}} A_\s\otimes\vartheta_{m+1}(B_0)\Omega},
\end{equation}
where we may assume that $A$ acts trivially on factors of $\h_\c$ with index $>N$. As in Section \ref{msect2} we replace the product of the propagators in \fer{m65} by $\e^{\i\tau K_1}\cdots\e^{\i\tau K_N} M^{m-N-1}$. Taking the limit $m=m(t)\rightarrow\infty$ then yields
\begin{equation}
\lim_{t\rightarrow\infty} \left| \,\C(t;A,A_\s,B_0) - \scalprod{\Omega}{(B')^*B' A\Omega}\ \omega_+\big( P\alpha_{\rm RI}^{s(t)}(A_\s\otimes B_0)P\big)\right|=0,
\label{m66}
\end{equation}
with a speed of convergence dictated by Proposition \ref{proposition3}. To complete the proof of Theorem \ref{thm3} we notice that $\scalprod{\Omega}{(B')^*B'A \Omega} = \av{A}_{B'\Omega} = \omega(A) +O(\varepsilon)$, for arbitrary $\varepsilon>0$ (c.f. the argument before \fer{m51}). \hfill $\blacksquare$

\subsection{Proof of Proposition \ref{proposition1}}
\label{proofprop1}
Let $K_m$, $L_m$ be the operators on $\h$ that act trivially on all factors except on $\h_\s$ times the $m$--th $\h_\ee$, where they act as $K$, $L$. Given any $t\in \r$, $A, B\in\mm_\s$ ($\subset \mm$) we have 
\begin{eqnarray}
\lefteqn{\scalprod{B\Omega_\s\otimes\Omega_\c}{\e^{\i tL_1}\cdots \e^{\i t L_m} A  \e^{-\i tL_m}\cdots \e^{-\i t L_1}\Omega_\s\otimes\Omega_\c}_{\h}}\nonumber\\
 &=& \scalprod{B\Omega_\s\otimes\Omega_\c}{\e^{\i tK_1}\cdots \e^{\i t K_m} A\Omega_\s\otimes\Omega_\c}_{\h}\nonumber\\
&=&\scalprod{B\Omega_\s}{(P\e^{\i tK}P)^m A\Omega_\s}_{\h_\s},
\label{m18}
\end{eqnarray}
where we use that $K$ and $L$ implement the same dynamics on $\mm$ and \fer{m15} in the first step, and Proposition \ref{proposition2} in the second step. Since $B\Omega_\s$ is dense in $\h_\s$ it follows from \fer{m18} that
\begin{equation}
\|(P \e^{\i tK}P)^m A\Omega_\s\|\leq \|A\|.
\label{m19}
\end{equation}
So far we have not used that $\dim \h_\s<\infty$. In the finite-dimensional case, $\mm_\s\Omega_\s$ is not only dense in $\h_\s$, but for any $\psi\in\h_\s$ there exists an $A\in\mm_\s$ s.t. $\psi=A\Omega_\s$. Thus \fer{m19} and the uniform boundedness principle give that 
\begin{equation}
\sup_{t\in\r, m\geq 0}\| (P\e^{\i tK}P)^m\|<\infty.
\label{m20}
\end{equation}
The facts that the spectrum of $P\e^{\i t K}P$ lies in the unit disk in $\cx$, and that all eigenvalues on the unit circle must be semisimple follow from the uniform boundedness of $\|(P\e^{\i tK}P)^m\|$ in $m$. They can be shown using an easy Jordan canonical form argument. \hfill $\blacksquare$

\subsection{Proof of Proposition \ref{proposition4}}
\label{sectprop4}

Given any normal state $\omega$ of $\mm$ and any unitary $U$ on $\h$ we have the following relation \cite{JP1} (our definition of entropy differs from the one in \cite{JP1} by a sign)
\begin{eqnarray}
\lefteqn{
{\rm Ent}\big(\omega(U^*\cdot U)|\omega_0\big) - {\rm Ent}(\omega| \omega_0)}\nonumber\\
&& =  \omega\Big(U^*\Big[\beta_\ee\sum_k L_{\ee,k} +\beta_\s L_\s\Big]U   -\beta_\ee\sum_k L_{\ee,k} -\beta_\s L_\s \Big).
\label{m74}
\end{eqnarray}
In case $U$ is a dynamics of the system, \fer{m74} is called the {\it entropy production formula}. We take $U=U_{\rm RI}(t)$. The argument of $\omega$ in \fer{m74} can be written as
\begin{equation*}
\beta_\ee \Big\{ \alpha_{\rm RI}^t\Big( \sum_k L_{\ee,k} +L_\s\Big) -\sum_k L_{\ee,k} -L_\s\Big\}-(\beta_\ee-\beta_\s)\Big( \alpha_{\rm RI}^t (L_\s )-L_\s\Big),
\end{equation*}
so it suffices to prove \fer{m75} for $\beta_\s=\beta_\ee=\beta$. We want to show that
\begin{equation}
\alpha_{\rm RI}^t\Big( \sum_k L_{\ee,k} +L_\s\Big) -\sum_k L_{\ee,k} -L_\s
= \Delta E(t) -\alpha_{\rm RI}^t(V_{m(t)+1}) +V_1.
\label{m84}
\end{equation}
It is clear that the sums in the l.h.s. of \fer{m84} extend only from $k=1$ to $k=m+1$. We examine the difference of the first two terms on the l.h.s., for $k=m+1$. With $t=m\tau+s$, we obtain for this difference the expression
\begin{eqnarray}
\lefteqn{
\alpha_{\rm RI}^t(L_{\ee,m+1}+L_\s+V_{m+1})-\alpha_{\rm RI}^t(V_{m+1})-L_{\ee,m+1}}\nonumber\\
&=& \alpha_{\rm RI}^{m\tau}(L_{\ee,m+1}+L_\s+V_{m+1})-\alpha_{\rm RI}^t(V_{m+1})-L_{\ee,m+1}\nonumber\\
&=& \alpha_{\rm RI}^{m\tau}(L_\s) +\alpha_{\rm RI}^{m\tau}(V_{m+1})-\alpha_{\rm RI}^t(V_{m+1})\nonumber\\
&=& \alpha_{\rm RI}^{m\tau}(L_\s+V_m) +\alpha_{\rm RI}^{m\tau}(V_{m+1}-V_m)-\alpha_{\rm RI}^t(V_{m+1})\nonumber\\
&=& \alpha_{\rm RI}^{m\tau}(L_\s+V_m) +j(m)-\alpha_{\rm RI}^t(V_{m+1}),
\label{m85}
\end{eqnarray}
where we use in the second step $\alpha_{\rm RI}^{m\tau}(L_{\ee,m+1}) =L_{\ee,m+1}$, and in the last step we use definition \fer{m68}. We now add to \fer{m85} the expression $\alpha_{\rm RI}^t(L_{\ee,m})-L_{\ee,m} = \alpha_{\rm RI}^{m\tau}(L_{\ee,m})-L_{\ee,m} $ (i.e., the term with $k=m$ in the sums of the l.h.s. of \fer{m84}) and repeat the manipulations leading to \fer{m85} to obtain for this sum the expression
\begin{equation*}
\alpha_{\rm RI}^{(m-1)\tau}(L_\s+V_{m-1}) +j(m-1)+j(m) - \alpha_{\rm RI}^t(V_{m+1}).
\end{equation*}
It is now clear how to continue this process until all terms with $k=1,\ldots,m+1$ in the sums of the l.h.s. of \fer{m84} are taken care of. We obtain
\begin{equation}
{\rm l.h.s.\  of \ \fer{m84}} =\sum_{k=1}^{m(t)} j(k) +V_1 -\alpha_{\rm RI}^t(V_{m(t)+1}).
\label{m86}
\end{equation}
Definition \fer{m72} yields the result \fer{m75}. 

To see that ${\rm Ent}\big(\omega\circ\alpha_{\rm RI}^t|\omega_0\big)$ is continuous in $t\geq 0$ we show that
\begin{equation}
{\rm Ent}\big(\omega\circ\alpha_{\rm RI}^{m\tau +s}|\omega_0\big) - {\rm Ent}\big(\omega\circ\alpha_{\rm RI}^{(m-1)\tau+s'}|\omega_0\big)
\label{m87}
\end{equation}
converges to zero, for all $m\geq 1$, as $s\downarrow 0$ and $s'\uparrow\tau$. Expression \fer{m75} yields
\begin{eqnarray}
\fer{m87} &=& \omega\Big( \beta_\ee\big[ \Delta E(m\tau+s) -\Delta E((m-1)\tau+s')\big]\label{m88}\\
&& \ \ \ \ \ -\alpha_{\rm RI}^{m\tau +s}\big(\beta_\ee V_{m+1}+\Delta\beta L_{\cal S}\big) +\alpha_{\rm RI}^{(m-1)\tau+s'}\big(\beta_\ee V_m+\Delta\beta L_{\cal S}\big)\Big),
\nonumber
\end{eqnarray}
where $\Delta \beta=\beta_\ee-\beta_{\cal S}$. The first argument of $\omega$ in \fer{m88} equals $\beta_\ee \,j(m)$, see \fer{m72}. Next, 
\begin{equation}
\alpha_{\rm RI}^{m\tau+s}(V_{m+1}) -\alpha_{\rm RI}^{(m-1)\tau+s'}(V_m) \longrightarrow \alpha_{\rm RI}^{m\tau}(V_{m+1}-V_m),
\label{m89}
\end{equation}
as $s\downarrow 0$, $s'\uparrow\tau$. But the r.h.s. of \fer{m89} is just $j(m)$, see \fer{m68}. This will compensate the contribution of the first argument in $\omega$ of \fer{m88} in the limit. Finally, $\alpha_{\rm RI}^{m\tau+s}(L_{\cal S}) -\alpha_{\rm RI}^{(m-1)\tau+s'}(L_{\cal S})$ tends to zero, as $s\downarrow0$, $s'\uparrow\tau$. Consequently, \fer{m87} vanishes in the limit. \hfill $\blacksquare$

\subsection{Proof of Proposition \ref{prop5}}

By \fer{m75} and \fer{m72}, the change of entropy within an interval of duration $\tau$ is given by 
\begin{eqnarray}
\lefteqn{
{\rm Ent}(\omega\circ\alpha_{\rm RI}^{t+\tau}|\omega_0)- {\rm Ent}(\omega\circ\alpha_{\rm RI}^t|\omega_0)}\label{b2}\\
&=&
\beta_\ee\,\omega\big( j(m+1) -\alpha_{\rm RI}^{t+\tau}(V_{m+2})+\alpha_{\rm RI}^t(V_{m+1}) \big) \label{m78.21}\\
&&-(\beta_\ee-\beta_\s)\,\omega\big( \alpha_{\rm RI}^{t+\tau}(L_\s)-\alpha_{\rm RI}^t(L_\s)\big),
\label{m78.2}
\end{eqnarray}
where $j(k)$ is given in \fer{m68}, and $m=m(t)$. Taking into account \fer{m68}, \fer{m70} and Theorem \ref{thm1} we see that the limit $t\rightarrow\infty$ of \fer{m78.21} is $\beta_\ee\,\omega_+(j_+)$. We claim that \fer{m78.2} vanishes as $t\rightarrow\infty$. This can be seen in the following way. An application of Theorem \ref{thm2} shows that 
\begin{equation}
\lim_{t\rightarrow\infty} \omega\big( \alpha_{\rm RI}^{t+\tau}(L_\s)-\alpha_{\rm RI}^t(L_\s)\big) = \int_0^\tau\omega_+(P\alpha_{\rm RI}^s(\i[V,L_\s])P)\d s,
\label{b1}
\end{equation}
where we use that $[V,L_\s]\in\mm_\s\otimes\mm_\ee$, which follows from the fact that $\e^{\i tL_\s}V\e^{-\i tL_\s}\in \mm_\s\otimes\mm_\ee$ for all $t\in\r$. Therefore
\begin{equation}
{\rm Ent}(\omega\circ\alpha_{\rm RI}^{t+\tau}|\omega_0)- {\rm Ent}(\omega\circ\alpha_{\rm RI}^t|\omega_0)\longrightarrow \beta_\ee\, \omega_+(j_+)-(\beta_\ee-\beta_\s)\int_0^\tau\!\!\omega_+(P\alpha_{\rm RI}^s(\i[V,L_\s])P)\d s
\label{b3}
\end{equation}
as $t\rightarrow\infty$. On the other hand, \fer{m76} shows that 
\begin{equation}
\frac 1t \big[{\rm Ent}(\omega\circ\alpha_{\rm RI}^{t+\tau}|\omega_0)- {\rm Ent}(\omega\circ\alpha_{\rm RI}^t|\omega_0)\big] \longrightarrow \frac{\beta_\ee}{\tau}\omega_+(j_+),
\label{b4}
\end{equation}
in the limit $t\rightarrow\infty$. Then the following general fact proves that \fer{b1} must be equal to zero: 

If a locally bounded function $f$ on $\r$ has the property $f(t+\tau)-f(t)\rightarrow a$, as $t\rightarrow\infty$, for some $a\in\r$ and some $\tau>0$, then $f(t)/t\rightarrow a/\tau$, as $t\rightarrow\infty$.

This concludes the proof of Proposition \fer{prop5}. \hfill $\blacksquare$

\subsection{Proof of Theorem \ref{thm:ergodic1}}\label{sec:prooferg1}

Using a Dyson expansion we get
\begin{eqnarray}\label{dyson}
\e^{\i \tau K_\lambda} & = & \e^{\i \tau K_0}+\i\lambda\int_0^\tau \d t \e^{\i(\tau-t)K_0}W\e^{\i tK_0}\\
 & & \qquad \qquad -\lambda^2\int_0^\tau\!\int_0^t \e^{\i(\tau-t)K_0}W\e^{\i(t-s)K_0}W\e^{\i sK_0}\,\d s \,\d t+O(\lambda^3).\nonumber
\end{eqnarray}
Inserting \fer{systliouv1}-\fer{chainliouv1}-\fer{interaction1}-\fer{cliouvint1} in \fer{dyson}, one gets after a somewhat lengthy but straightforward computation
\begin{eqnarray}\label{Mpert1}
M_\lambda & = & \e^{\i\tau L_\s}+\i\lambda \sin(\tau)\left\|\e^{-\beta h/2}g_\beta\right\|_\fh^2 \left(\sigma_x\otimes\e^{-\i\tau\sigma_z}-\e^{\i\tau\sigma_z}\otimes\sigma_x\right)\nonumber\\
 & & +\lambda^2\int_0^\tau \!\int_0^t\left(\int_{\R^+}\!\!\int_{\R^+}\|g_\beta(r_1)\|^2\|g_\beta(r_2)\|^2\e^{-\beta r_1}\e^{\i(t-s)(r_2-r_1)}\d r_1\,\d r_2\right. \nonumber\\
 & & \qquad \qquad \qquad  +\|\e^{-\beta h/2}g_\beta\|_\fh^4\Big)\left(\e^{\i(\tau-t)\sigma_z}\sigma_x \e^{\i t\sigma_z}\otimes \e^{-\i(\tau-s)\sigma_z}\sigma_x \e^{-\i s\sigma_z}\right.\nonumber\\
 & & \qquad \qquad \qquad \qquad \qquad \qquad \qquad \qquad \left. -\e^{\i(\tau-2t+2s)\sigma_z}\otimes\e^{-\i\tau\sigma_z}\right)\d s\,\d t \nonumber\\
 & & +\lambda^2\int_0^\tau \!\int_0^t\left(\int_{\R^+}\!\!\int_{\R^+}\|g_\beta(r_1)\|^2\|g_\beta(r_2)\|^2\e^{-\beta r_2}\e^{\i(t-s)(r_2-r_1)}\d r_1\,\d r_2\right. \nonumber\\
 & & \qquad \qquad \qquad  +\|\e^{-\beta h/2}g_\beta\|_\fh^4\Big)\left(\e^{\i(\tau-s)\sigma_z}\sigma_x \e^{\i s\sigma_z}\otimes \e^{-\i(\tau-t)\sigma_z}\sigma_x \e^{-\i t\sigma_z}\right.\nonumber\\
 & & \qquad \qquad \qquad \qquad \qquad \qquad \qquad \qquad \left. -\e^{\i\tau\sigma_z}\otimes\e^{-\i(\tau-2t+2s)\sigma_z}\right)\d s\,\d t \nonumber\\
 & & +O(\lambda^3).
\end{eqnarray}
Using perturbation theory \cite{K}, we then find that $M_\lambda$ has
$4$ distinct eigenvalues: 
$1$, $e_0(\lambda)$, $e_+(\lambda)$, $e_-(\lambda)$ which are given by
\beq
e_0(\lambda)=1-\lambda^2\tau^2(\alpha_1+\alpha_2)+O(\lambda^3), \qquad \qquad \qquad \qquad \qquad \qquad \qquad \label{evalue01}
\eeq
\begin{eqnarray}\label{evalue+1}
e_+(\lambda) & = & \e^{2\i\tau}\left[1-\frac{\lambda^2\tau^2}{2}(\alpha_1+\alpha_2) +\i\lambda^2\tau\Big( \|\e^{-\beta h/2}g_\beta\|_\fh^4 \right.\\
 & & \qquad \qquad \qquad -\int\!\!\int (\e^{-\beta r_1}+\e^{-\beta r_2})\|g_\beta(r_1)\|_\fg^2\|g_\beta(r_2)\|_\fg^2\nonumber \\
 & & \qquad \qquad \qquad \qquad \qquad \left. \times \frac{1-\sinc(\tau(r_1-r_2-2))}{r_1-r_2-2}\Big) \right]+O(\lambda^3),\nonumber
\end{eqnarray}
\begin{eqnarray}\label{evalue-1}
e_-(\lambda) & = & \e^{-2\i\tau}\left[1-\frac{\lambda^2\tau^2}{2}(\alpha_1+\alpha_2) -\i\lambda^2\tau\Big( \|\e^{-\beta h/2}g_\beta\|_\fh^4 \right.\\
 & & \qquad \qquad \qquad -\int\!\!\int (\e^{-\beta r_1}+\e^{-\beta r_2})\|g_\beta(r_1)\|_\fg^2\|g_\beta(r_2)\|_\fg^2\nonumber \\
 & & \qquad \qquad\qquad \qquad \qquad \left. \times \frac{1-\sinc(\tau(r_1-r_2-2))}{r_1-r_2-2}\Big) \right]+O(\lambda^3),\nonumber
\end{eqnarray}
where the $\alpha_j$'s are defined in \fer{alphaj1}. Since they are strictly positive numbers, this proves that for $|\lambda|$ small enough, the operator $M_\lambda$ satisfies Assumption (E).

It remains to prove that the asymptotic state $\omega_{+,\lambda}$ is indeed given by \fer{asympstate1}. For that purpose, it suffices to compute the eigenvector $\Omega_\s^*(\lambda)$ of $M_\lambda^*$ for the eigenvalue $1,$ which is obtained by perturbation theory: 
\begin{eqnarray}\label{omega*1}
\Omega_\s^*(\lambda) & = & \frac{\alpha_1\sqrt{2}}{\alpha_1+\alpha_2}\psi_{11}+\frac{\alpha_2\sqrt{2}}{\alpha_1+\alpha_2}\psi_{22}\\
 & & \qquad \qquad +\lambda\|\e^{-\beta h/2}g_\beta\|^2\frac{\alpha_1-\alpha_2}{\sqrt{2}(\alpha_1+\alpha_2)}(\psi_{12}+\psi_{21})+O(\lambda^2).\nonumber
\end{eqnarray}

%%%%%%%%%%%%%%%%%%%%%%%%%%%%%%%%%%%%%%%%%%%%%%%%%%%%%%%%%%%%%%%%%%%%%%%%%%%%%%%%%%%%%
%%%%%%%%%%%%%%%%%%%%%%%%%%%%%%%%%%%%%%%%%%%%%%%%%%%%%%%%%%%%%%%%%%%%%%%%%%%%%%%%%%%%%

\subsection{Proof of Theorem \ref{thm:entropy1}}\label{sec:proofent1}

Let $P_0$ denote the spectral projection on the kernel of $L_0$, and
$\Omega_0^*$ the main term in the expansion of $\Omega_\s^*(\lambda)$,
i.e. $\Omega_\s^*(\lambda)=\Omega_0^*+O(\lambda)$ (see
e.g. \fer{omega*1}). Then, using \fer{m22}, \fer{m70} and perturbation theory, one gets 
\begin{eqnarray}\label{entrop-pert}
\omega_+(j_+) & = & \lambda^2\left[\i\int_0^\tau \langle
  \Omega^*_0\otimes\Omega_\ee |V(\e^{-\i tL_0}-\e^{\i
    tL_0})(1-P)V\Omega_\s\otimes\Omega_\ee\rangle \right.\nonumber \\
 & & \qquad \qquad \left. -\i\int_0^\tau
  \langle \Omega_0^*\otimes\Omega_\ee | WP_0\e^{\i tL_0} V \Omega_\s\otimes\Omega_\ee\rangle\right]  +O(\lambda^3).
\end{eqnarray}
Moreover, as a general fact, one has $P_0
\Omega_0^*\otimes\Omega_\ee=\Omega_0^*\otimes\Omega_\ee$. Now, the
quadratic interaction satisfies $P_0WP_0=0.$ Hence the second term on the right hand side of
\fer{entrop-pert} cancels. 

Using \fer{m81},\fer{interaction1} and \fer{omega*1} we thus have
\begin{eqnarray}
\d S_+ & = & \frac{\lambda^2\beta\tau}{\alpha_1+\alpha_2}\int \|g_\beta(r)\|^2\|g_\beta(r')\|^2(2-r+r')\sinc^2\left(\frac{\tau(2-r+r')}{2}\right)\nonumber\\
 & & \qquad\qquad\qquad\qquad\qquad\qquad  \times(\alpha_2\e^{-\beta r}-\alpha_1\e^{-\beta r'})\d r\d r'\,+O(\lambda^3).\nonumber
\end{eqnarray}
Inserting the expression for $\alpha_j$ in the integral, one finally has
\begin{eqnarray}\label{entropy1}
\d S_+ & = & \frac{\lambda^2\beta\tau}{2(\alpha_1+\alpha_2)}\int \|g_\beta(r_1)\|^2\|g_\beta(r_2)\|^2\|g_\beta(r_3)\|^2\|g_\beta(r_4)\|^2\\
 & & \qquad \qquad \times \sinc^2\left(\frac{\tau(2-r_1+r_2)}{2}\right)\sinc^2\left(\frac{\tau(2-r_3+r_4)}{2}\right)\nonumber \\
 & & \qquad \qquad \times (r_2+r_3-r_1-r_4)\left(\e^{-\beta(r_1+r_4)}-\e^{-\beta(r_2+r_3)}\right)+O(\lambda^3),\nonumber
\end{eqnarray}
where the integral runs over the four variables $r_j\in\R^+.$ The result follows then from the fact that, for any real $x$ and $y,$ $(x-y)(\e^{-\beta y}-\e^{-\beta x})$ is non-negative.\hfill\qed

%%%%%%%%%%%%%%%%%%%%%%%%%%%%%%%%%%%%%%%%%%%%%%%%%%%%%%%%%%%%%%%%%%%%%%%%%%%%%%%%%%%%%
%%%%%%%%%%%%%%%%%%%%%%%%%%%%%%%%%%%%%%%%%%%%%%%%%%%%%%%%%%%%%%%%%%%%%%%%%%%%%%%%%%%%%

\subsection{Proof of Theorem \ref{thm:ergodic2}}\label{sec:prooferg2}

The proof goes exactly in the same way as for Theorems \ref{thm:ergodic1} and \ref{thm:entropy1}. We just give the expressions for the eigenvalues (different from $1$) of $M_\lambda$ as well as the one for the entropy production.
\beq\label{evalue02}
e_0(\lambda)=1-\lambda^2\tau^2(\alpha_1+\alpha_2)+O(\lambda^4),\qquad\qquad\qquad\qquad\qquad\qquad\qquad\qquad
\eeq
\begin{eqnarray}\label{evalue+2}
e_+(\lambda) & = & \e^{2\i\tau}\left[ 1-\frac{\lambda^2\tau^2}{2}(\alpha_1+\alpha_2) +\i\lambda^2\tau^2\int \|g(r)\|_\fg^2\right.\\
 & & \qquad \quad \left. \times \left(\frac{1-\sinc(\tau(2-r))}{2-r}+\frac{1-\sinc(\tau(2+r))}{2+r}\right)\right]+O(\lambda^4),\nonumber
\end{eqnarray}
\begin{eqnarray}\label{evalue-2}
e_+(\lambda) & = & \e^{-2\i\tau}\left[ 1-\frac{\lambda^2\tau^2}{2}(\alpha_1+\alpha_2) -\i\lambda^2\tau^2\int \|g(r)\|_\fg^2\right.\\
 & & \qquad \quad \left. \times \left(\frac{1-\sinc(\tau(2-r))}{2-r}+\frac{1-\sinc(\tau(2+r))}{2+r}\right)\right]+O(\lambda^4),\nonumber
\end{eqnarray}
\begin{eqnarray}\label{entropy2}
\d S_+ & = & \frac{\lambda^2\beta\tau}{\alpha_1+\alpha_2}\int \|g_\beta(r)\|_\fg^2\,\|g_\beta(r')\|_\fg^2\,\sinc^2\left(\frac{\tau(r-2)}{2}\right)\\
 & & \qquad \qquad\qquad\qquad \sinc^2\left(\frac{\tau(r'+2)}{2}\right)(r+r')(1-\e^{-\beta(r+r')})\nonumber \\
 & & +\frac{\lambda^2\beta\tau}{2(\alpha_1+\alpha_2)}\int \|g_\beta(r)\|_\fg^2\,\|g_\beta(r')\|_\fg^2\, (r'-r)(\e^{-\beta r}-\e^{-\beta r'})\nonumber\\
 & & \qquad \qquad  \times\left[\sinc^2\left(\frac{\tau(r-2)}{2}\right)\sinc^2\left(\frac{\tau(r'-2)}{2}\right)\right. \nonumber \\
 & & \quad \qquad \qquad \left. +\sinc^2\left(\frac{\tau(r+2)}{2}\right)\sinc^2\left(\frac{\tau(r'+2)}{2}\right)\right]+O(\lambda^3).\nonumber
\end{eqnarray}
All the integrals run over $\R^+$, and in the formula for the entropy the integration is computed w.r.t. both $r$ and $r'$.
\hfill\qed

%%%%%%%%%%%%%%%%%%%%%%%%%%%%%%%%%%%%%%%%%%%%%%%%%%%%%%%%%%%%%%%%%%%%%%%%%%%%%%%%%%%%%
%%%%%%%%%%%%%%%%%%%%%%%%%%%%%%%%%%%%%%%%%%%%%%%%%%%%%%%%%%%%%%%%%%%%%%%%%%%%%%%%%%%%%

\subsection{Proof of Theorem \ref{thm:ergodic3}}\label{sec:prooferg3}

Once again the proof goes the same way, and we only give the expressions for the eigenvalues (different from $1$) of $M_\lambda$ and for the entropy production.
\beq\label{evalue03}
e_0(\lambda)=1-\frac{\lambda^2\tau^2}{1+\e^{-\beta E_\ee}}(\alpha_1+\alpha_2)+O(\lambda^4),\qquad\qquad\qquad\qquad\qquad\qquad\qquad\qquad\nonumber
\eeq

\begin{eqnarray}\label{evalue+3}
e_+(\lambda) & = & \e^{\i\tau E_\s}\left[1-\frac{\lambda^2\tau^2}{2(1+\e^{-\beta E_\ee})}\Big(\alpha_1+\alpha_2+ (1+\e^{-\beta E_\ee})\sinc^2\left(\frac{\tau E_\ee}{2}\right)\right.\nonumber \\ 
 & & \qquad \qquad \qquad \qquad \qquad \qquad \qquad \qquad  \times(|\eba|^2+|\ebd|^2-\bar{\eba}\ebd-\eba\bar{\ebd})\Big)\nonumber \\
 & & \qquad \qquad  +\i\frac{\lambda^2\tau^2}{1+\e^{-\beta E_\ee}}\left( (1-\e^{-\beta E_\ee})\frac{1-\sinc(\tau E_\ee)}{\tau E_\ee}(|\eba|^2-|\ebd|^2)\right.\nonumber\\
 & & \qquad \qquad \qquad \qquad \qquad +(1-\e^{-\beta E_\ee})\sinc^2\left(\frac{\tau E_\ee}{2}\right)\frac{\bar{\eba}\ebd-\eba\bar{\ebd}}{2\i}\nonumber\\
 & & \qquad \qquad \qquad \qquad \qquad  -(1+\e^{-\beta E_\ee})\frac{1-\sinc(\tau(E_\ee-E_\s))}{\tau(E_\ee-E_\s)}|\ebb|^2\nonumber\\
 & & \qquad \qquad \qquad \qquad \qquad \left. \left. +(1+\e^{-\beta E_\ee})\frac{1-\sinc(\tau(E_\ee+E_\s))}{\tau(E_\ee+E_\s)}|\ebc|^2 \right) \right]\nonumber\\
 & &  +O(\lambda^4),\nonumber
\end{eqnarray}

\begin{eqnarray}\label{evalue-3}
e_-(\lambda) & = & \e^{-\i\tau E_\s}\left[1-\frac{\lambda^2\tau^2}{2(1+\e^{-\beta E_\ee})}\Big(\alpha_1+\alpha_2+ (1+\e^{-\beta E_\ee})\sinc^2\left(\frac{\tau E_\ee}{2}\right)\right.\nonumber \\ 
 & & \qquad \qquad \qquad \qquad \qquad \qquad \qquad \qquad  \times(|\eba|^2+|\ebd|^2-\bar{\eba}\ebd-\eba\bar{\ebd})\Big)\nonumber \\
 & & \qquad \qquad  +\i\frac{\lambda^2\tau^2}{1+\e^{-\beta E_\ee}}\left( (1-\e^{-\beta E_\ee})\frac{1-\sinc(\tau E_\ee)}{\tau E_\ee}(|\ebd|^2-|\eba|^2)\right.\nonumber\\
 & & \qquad \qquad \qquad \qquad \qquad +(1-\e^{-\beta E_\ee})\sinc^2\left(\frac{\tau E_\ee}{2}\right)\frac{\eba\bar{\ebd}-\bar{\eba}\ebd}{2\i}\nonumber\\
 & & \qquad \qquad \qquad \qquad \qquad  +(1+\e^{-\beta E_\ee})\frac{1-\sinc(\tau(E_\ee-E_\s))}{\tau(E_\ee-E_\s)}|\ebb|^2\nonumber\\
 & & \qquad \qquad \qquad \qquad \qquad \left. \left. -(1+\e^{-\beta E_\ee})\frac{1-\sinc(\tau(E_\ee+E_\s))}{\tau(E_\ee+E_\s)}|\ebc|^2 \right) \right]\nonumber\\
 & &  +O(\lambda^4),\nonumber
\end{eqnarray}

\begin{eqnarray}\label{entropy3}
\d S_+ & = & \frac{\lambda^2\beta\tau E_\ee(1-\e^{-\beta E_\ee})}{(\alpha_1+\alpha_2)(1+\e^{-\beta E_\ee})}\times\nonumber\\ 
 & & \qquad  \left[ |\ebb|^2(|\eba|^2+\e^{-\beta E_\ee}|\ebd|^2)\sinc^2\left(\frac{\tau(E_\ee-E_\s)}{2}\right)\sinc^2\left(\frac{\tau E_\ee}{2}\right)\right. \nonumber\\
 & & \qquad +|\ebc|^2(\e^{-\beta E_\ee}|\eba|^2+|\ebd|^2)\sinc^2\left(\frac{\tau(E_\ee+E_\s)}{2}\right)\sinc^2\left(\frac{\tau E_\ee}{2}\right)\nonumber\\
 & & \qquad  +2|\ebb|^2|\ebc|^2(1+\e^{-\beta E_\ee})\sinc^2\left(\frac{\tau(E_\ee+E_\s)}{2}\right) \nonumber\\ 
 & & \qquad\qquad\qquad\qquad\qquad\qquad \left. \times\,\sinc^2\left(\frac{\tau (E_\ee+E_\s)}{2}\right)\right] +O(\lambda^3).\nonumber
\end{eqnarray}
\hfill\qed

\bigskip
{\bf Acknowledgements.\ } This work was initiated during a visit of
M.M. to the Institut Fourier of Universit\'e de Grenoble I while
L.B. was a postdoc there, and it was essentially completed during a visit of L.B. to the Mathematics Department of McGill University. We are grateful to the Institut Fourier for financial support and to both  the Institut Fourier and McGill University for their hospitality.


\begin{thebibliography}{99}

\bibitem{AW} 
Araki, H., Wyss, W.: {\it Representations of canonical
        anticommutation relations} Helv. Phys. Acta {\bf 37} 136--159
      (1964).

\bibitem{AJ} 
Attal, S., Joye, A.: {\it Weak Coupling and Continuous
        Limits for Repeated Quantum Interactions} preprint mp-arc 05-5.

\bibitem{AP}
Attal, S., Pautrat, Y.: {\it From repeated to continuous quantum interactions.}  To appear.

\bibitem{WBKM}
Wellens, T., Buchleitner, A., K\"ummerer, B., Maassen, H.: {\it Quantum State Preparation via Asymptotic Completeness.}  Phys. Rev. Lett. {\bf 85}, no.16, 3361--3364 (2000)


\bibitem{BR}
Bratteli, O., Robinson, D.W., Operator Algebras and Quantum Statistical Mechanics I, II. Texts and Monographs in Physics, Springer-Verlag, 1987.


\bibitem{JP}
Jak\u{s}i\'c, V., Pillet, C.-A.: {\it Non-equilibrium steady states of finite quantum systems coupled to thermal reservoirs.}  Comm. Math. Phys.  {\bf 226}, no.1, 131--162  (2002).


\bibitem{JP1}
Jak\u{s}i\'c, V., Pillet, C.-A.: {\it A note on the entropy production formula.}  Advances in differential equations and mathematical physics (Birmingham, AL, 2002),  175--180, Contemp. Math., 327, Amer. Math. Soc., Providence, RI, 2003.

\bibitem{K} Kato, K., Perturbation Theory for Linear Operators. $2nd$
      edition. Springer, Berlin, 1976.


\bibitem{MMS}
Merkli, M., M\"uck, M., Sigal, I.M.: {\it Instability of Equilibrium States for Coupled Heat Reservoirs at Different Temperatures.}\ submitted.


\bibitem{P} 
Pillet, C.-A., {\it Quantum Dynamical Systems.} Springer Lecture Notes in Mathematics, to appear


\bibitem{VAS}
Vogel, K., Akulin, V.M., Schleich, W.P.: {\it Quantum State Engineering of the Radiation Field.} Phys. Rev. Lett. {\bf 71}, no.12, 1816--1819 (1993)


\end{thebibliography}
\end{document}